\documentclass[11pt]{article}      
\pdfoutput=1
\usepackage{amsthm,amsmath,amssymb,xcolor,sectsty,latexsym,mathtools,hyperref,mathrsfs}
\usepackage{slashed}  
\usepackage{bm}
\usepackage{epsfig}
\usepackage{dcolumn}
\definecolor{vermelho}{cmyk}{0,.88,.77,.40}
\usepackage{cite}
\usepackage{feynmp}
\numberwithin{equation}{section}
\usepackage[textwidth=6.5in,top=1.2in,bottom=1.2in]{geometry}
\usepackage{setspace}
\chapterfont{\color{vermelho}}
\usepackage{simplewick}
\usepackage{hyperref}

\newcommand{\be}{\begin{equation}}
\newcommand{\ee}{\end{equation}}
\newcommand{\beq}{\begin{equation}}
\newcommand{\eeq}{\end{equation}}
\newcommand{\ba}{\begin{eqnarray}}
\newcommand{\ea}{\end{eqnarray}}
\newcommand{\bef}{\begin{figure}}
\newcommand{\eef}{\end{figure}}

\newcommand{\tr}{\xi}
\newcommand{\trr}{\bar\xi}
\newcommand{\Con}{\Lambda}
\newcommand{\eps}{\gamma}
\newcommand{\Neff}{N_{\tiny\mbox{eff}}}
\newcommand{\X}{\varphi}
\newcommand{\UV}{\Lambda_{UV}}
\newcommand{\aUV}{\alpha_{UV}}
\newcommand{\Lam}{M}
\newcommand{\Lamv}{\bar{M}}
\newcommand{\mpl}{M_{Pl}}
\newcommand{\Tvs}{T_{vs}}
\newcommand{\Td}{T_{dark}}
\newcommand{\rhovs}{\rho_{vs}}
\newcommand{\rhod}{\rho_{dark}}
\newcommand{\gvs}{g_{vs}}
\newcommand{\gd}{g_{dark}}

\newcommand{\higgs}{h}
\newcommand{\ad}{\alpha_d}
\newcommand{\Nd}{N_{d}}
\newcommand{\Md}{M_{d}}
\newcommand{\nf}{n_{d}}
\newcommand{\gamm}{\epsilon}
\newcommand{\numd}{u_n}
\newcommand{\F}{f}
\newcommand{\lammix}{\bar\lambda}

\begin{document}

\thispagestyle{empty}
\begin{titlepage}
\nopagebreak

\title{  \begin{center}\bf Dark Matter and Naturalness \end{center} }

\vfill
\author{Mark P.~Hertzberg$^{1}$\footnote{mark.hertzberg@tufts.edu}, ~ McCullen Sandora$^{2}$\footnote{mccullen.sandora@gmail.com}}
\date{ }

\maketitle

\begin{center}
	\vspace{-0.7cm}
	{\it  $^{1}$Institute of Cosmology, Department of Physics and Astronomy}\\
	{\it  Tufts University, Medford, MA 02155, USA}\\
	{\it  $^{2}$Center for Particle Cosmology, Department of Physics and Astronomy}\\
	{\it University of Pennsylvania, Philadelphia, PA 19104, USA}
	\end{center}
\bigskip
\begin{abstract}
The Standard Model of particle physics is governed by Poincar\'e symmetry, while all other symmetries, exact or approximate, are essentially dictated by theoretical consistency with the particle spectrum. On the other hand, many models of dark matter exist that rely upon the addition of new added global symmetries in order to stabilize the dark matter particle and/or achieve the correct abundance. In this work we begin a systematic exploration into truly natural models of dark matter, organized by only relativity and quantum mechanics, without the appeal to any additional global symmetries, no fine-tuning, and no small parameters. We begin by reviewing how singlet dark sectors based on spin 0 or spin ${1\over2}$ should readily decay, while pure strongly coupled spin 1 models have an overabundance problem. This inevitably leads us to construct chiral models with spin ${1\over2}$ particles charged under confining spin 1 particles. This leads to stable dark matter candidates that are analogs of baryons, with a confinement scale that can be naturally $\mathcal{O}(100)$TeV. This leads to the right freeze-out abundance by annihilating into massless unconfined dark fermions. The minimal model involves a dark copy of $SU(3)\times SU(2)$ with 1 generation of chiral dark quarks and leptons. The presence of massless dark leptons can potentially give rise to a somewhat large value of $\Delta N_{\text{eff}}$ during BBN. In order to not upset BBN one may either appeal to a large number of heavy degrees of freedom beyond the Standard Model, or to assume the dark sector has a lower reheat temperature than the visible sector, which is also natural in this framework. This reasoning provides a robust set of dark matter models that are entirely natural. Some are concrete realizations of the nightmare scenario in which dark matter may be very difficult to detect, which may impact future search techniques.
\end{abstract}

\end{titlepage}

\setcounter{page}{2}

\tableofcontents

\newpage

\section{Introduction}

Modern physics has achieved a very succinct description of every physical process we have yet encountered, consisting of the Standard Model, dark matter, dark energy, and an extended neutrino sector; along with the ideas of baryogenesis and inflation.  However, it is difficult to understand how this could be the entire description. In particular, there are several features present in the Standard Model + $\Lambda$CDM that strike one as somewhat puzzling, or somehow nongeneric: why is the weak scale so small compared to the Planck scale? Why is the vacuum energy so small? Why are there three generations? Why are there so many accidental symmetries?  Why do the parameter values suggest that the Higgs is marginally stable?  

Arguably the most ``natural" scenario is a theory in which all physical scales are comparable, say, the Planck scale, and all dimensionless couplings are $\mathcal{O}(1)$. The observed smallness of the weak scale and the vacuum energy appears as a major challenge to this point of view. However, it is at least plausible that these scales happen to be small, or ``fine-tuned", due to environmental selection effects; the idea that these scales need to be so small for life to exist (increasing the weak scale can ruin the stability of nuclei \cite{Agrawal:1997gf} and increasing the vacuum energy can ruin the formation of galaxies \cite{Weinberg:1987dv}).
On the other hand, such environmental selection effects may not be efficient in the dark sector, since we are evidently {\em not} built out of the dark sector's degrees of freedom. 

However, a dark sector (or ``hidden sector") may provide the dark matter of our universe, which will be the focus of this paper (e.g., see Refs.~\cite{Kaplan:2009ag,Cohen:2010kn,Shelton:2010ta,Cheung:2010gj,Das:2010ts,Foot:2014uba,Blinov:2012hq,Lonsdale:2014wwa,Buckley:2014fba,Elor:2015bho,Acharya:2016fge,Dienes:2016vei,Escudero:2017yia,Tsao:2017vtn}). So could it be that the parameters of the dark sector also conspire to be fine-tuned simply for life to exist; namely that the gravitational interaction between the dark and visible sectors (or other possible weak interactions) are not too large or small as to make it difficult for galaxies, stars, etc to form? We take the point of view that this might be possible, although the argument is somewhat less potent compared to varying parameters in the Standard Model which obviously have a dramatic effect on life. While some authors have argued that a moderate increase in dark matter can be harmful to life \cite{Linde:1991km,Wilczek:2004cr,Tegmark:2005dy}, others have argued that we can readily tolerate a few orders of magnitude increase in the dark matter abundance before dramatic environmental problems occur \cite{Hellerman:2005yi}. 

So the following scenario is at least plausible and will be pursued here: The parameters of nature are highly natural on average, perhaps in some landscape framework \cite{Susskind:2003kw,Grana:2005jc,Douglas:2006es}. In the visible sector, parameters may appear fine-tuned, but this is due to selection effects or some other dynamical reason, while in the dark sector the parameters really are highly natural (and environmental selection effects are only relevant in extreme corners of parameter space). This leads to the model building exercise of constructing entirely natural models of dark matter that achieve the correct relic abundance and fit with observations. We will in fact build models where the relic abundance varies with the fundamental couplings on a logarithmic scale due to the renormalization group flow of couplings, and hence only requiring $\mathcal{O}(1)$ inputs. If the parameters of nature are unique, this seems much more palatable with ideas on unification. Moreover, in a landscape scenario, this would be exponentially preferred over any other models where the parameters need to be fine-tuned.

In the literature, often the idea associated with ``natural" models involves the concept of ``technical naturalness", in which it is said to be technically natural for a parameter to be small if the theory acquires a symmetry in the limit in which the parameter is taken to zero \cite{tHooft:1979rat}. This is associated with parameters being stable against large radiative corrections. While this is an interesting model building idea, there is little evidence that this is any kind of principle in nature; modern physics is built only on Poincar\'e symmetry and the rules of quantum mechanics. The only other symmetries observed are those that are {\em dictated} by these underlying principles. For example, CPT invariance is a property of any local Lorentz invariant theory. Furthermore, the Standard Model gauge group $SU(3)\times SU(2)\times U(1)$ is comprised of two aspects: (i) a pure ``gauge symmetry" part which is in fact a mere redundancy and can be removed from the theory by gauge fixing, and (ii) a global symmetry sub-group, which is a real symmetry, but is not optional, it is required as the only consistent way to have interacting massless spin 1 particles that obey the underlying principles \cite{Weinberg:1964ew}. Furthermore, the $U(1)_{B-L}$ follows as an accidental symmetry. Finally, any symmetry that can be broken {\em is} broken: this includes chiral symmetry, C, CP, scale invariance, etc. 

Now it is well known that quantum gravity is thought to forbid {\em exact} global symmetries \cite{Kallosh:1995hi,Banks:2010zn,Harlow:2018tng}, while allowing for {\em approximate} global symmetries; but this is a relatively mild statement. The more interesting issue is the significance and accuracy of approximate global symmetries. Consider the following: while it would have been technically natural for all the fermions in the Standard Model to be extremely light, as the theory would acquire an {\em approximate} chiral symmetry in this regime, it did not do this; the top quark's Yukawa coupling is $\mathcal{O}(1)$, while the up and down quark masses may be light due to environmental effects \cite{Barr:2007rd}. In summary, there is no current evidence of nature choosing even approximate global symmetries as a principle.

Guided by this, we will not impose any unessential (even approximate) global symmetries in the dark sector either. We begin a systematic exploration into the consequences of this. So if a particle, such as a scalar or fermion, is allowed to have a huge mass near some fundamental scale, like the GUT or Planck scale, then we will give it such a huge mass. If a particle can decay due to some Lorentz invariant operator, then we will include such an operator. We will not ignore these possibilities by appealing to unessential new global symmetries to prevent these problems (also see Refs.~\cite{Cata:2014sta,Antipin:2015xia,Mambrini:2015sia}), only accidental symmetries may emerge.  Instead we will hunt for models that lead to acceptable dark matter models, while being entirely natural and having no unneeded features; we only demand that Lorentz invariance (which could itself be conceivably derivable \cite{Hertzberg:2017nzl}) and quantum mechanics are obeyed. We will find that these simple criteria are already powerfully constraining, limiting the initially vast space of possibilities to a narrow few. Our main findings are summarized in Table \ref{tableone}.

Our paper is organized as follows:
In Section \ref{Spins} we give an overview of the spins of particles we will consider.
In Section \ref{Singlets} we begin with singlet scalars and fermions, showing that they invariably decay.
In Section \ref{Production} we discuss dark and visible sector production after inflation.
In Section \ref{YangMills} we discuss the case of pure Yang-Mills.
In Section \ref{ZeroOne} we discuss matter with abelian charges.
In Section \ref{chiral} we develop much more promising models that involve chiral matter and confinement.
In Section \ref{OneGenModel} we analyze a particular model involving $SU(3)\times SU(2)$ with one generation of chiral quarks and leptons in more detail.
In Section \ref{BBN} we compute constraints from big bang nucleosynthesis.
Finally, in Section \ref{Conclusions} we conclude.

\begin{table}[tb]
	\vskip.4cm
	\begin{center}
		\begin{tabular}{|c|c|c|c|}
			\hline 
			\bf{Theory} & \bf{Particles} & \bf{Mass} & \bf{Consequences}\\
			\hline
			Singlet Fermions & Massive fermions & $m_D$ & Huge mass; rapid decay\\
			\hline
			Singlet Scalars & Massive scalars & $m_\varphi$ & Huge mass; rapid decay\\
			\hline
			$SU(\Nd)$ & Glueballs & $\sim\Con$ & Overabundance\\
			$(3\rightarrow2)$ & & & Too hot\\
			\hline
			$U(1)$+Matter & Dark photons & $0$ & Photon mixing\\
		        & Charged matter & $m_d$ & Huge mass; overabundance\\
			\hline
			$SU(\Nd)$+Matter & Glueballs & $\sim\Con$ & Overabundance\\
			& Dark baryons & $\sim\Con,\Nd\,m_d$ & Huge mass allowed\\
			\hline
			$SU(\Nd)\times U(1)$, $\nf=2$ & Dark photons & $\sim e'\Con$ & Photon mixing \\
			 & Dark pions & $\sim e'\,\Con$ & \\
			 & Dark baryons & $\sim \Con$ & \\
			\hline
			$SU(\Nd)\times SU(2)$, $N_g\geq1$ & Dark leptons & 0 & (Possible) shift in $\Neff$ \\
			& $W^{1,2,3}$ & $\sim g\,\Con$ &  \\
			& Dark baryons & $\sim\Con$ & \\
			& $4N_g^2-4$ pions & 0 & \\
			\hline
			SM w/o Higgs, $N_g\geq1$ & Dark photons & 0 & Photon mixing\\   
		        & Charged leptons & 0 & Massless charged particles\\    
			& $W^\pm, Z$ & $\sim g\,\Con$ & (Possible) shift in $\Neff$ \\
			& Dark neutron & $\sim\Con$ &\\
			& Dark proton & $\sim(1+\alpha')\Con$ &\\
			& $2N_g^2-2$ neutral pions & 0 &  \\
			& $2N_g^2-2$ charged pions & $\sim e'\,\Con$ & \\
			\hline
		\end{tabular}
	\end{center}
	\caption{Table of dark matter models and their properties that we discuss in this paper ($\nf$ is number of Weyl fermions, $N_g$ is number of generations), including estimates of natural masses in terms of the strong coupling scale $\Con$, and some of their (potentially problematic) consequences. The second to last model is arguably the most natural (with minimal choice $\Nd=3$, $N_g=1$.) For simplicity, we have suppressed the dependence on $\Nd$, $N_g$ in our simple scaling estimates for dynamically generated masses.}
	\label{tableone}
\end{table}

\section{Overview of Spins}\label{Spins}

In this work we will restrict our attention to models that can be described in terms of light degrees of freedom.  We will therefore not study macroscopic objects, such as primordial black holes, which have been studied elsewhere. Furthermore, we will focus on models that are described by an effective field theory that has a cutoff well above the mass of the dark matter particle(s). The rules of relativity and quantum mechanics then leave only 5 possibilities for the spin of the particles $s=0,\,{1\over2},\,1,\,{3\over2},\,2$, since it is thought there is no consistent effective field theory of particles of spin $s>2$ with a high cutoff. Furthermore, we know $s=2$ is the graviton, which is definitely not the dark matter (we assume a massless graviton here). Also the case $s={3\over2}$ requires the introduction of supergravity, which is an interesting possibility, but we will not pursue this subject here (in any case, one would need to explain why it has a relatively low breaking scale for it to be relevant here). So our focus here will be on the only remaining possibilities for the spins of particles in the dark sector, namely
\beq
s=0,\,{1\over2},\,1.
\eeq 
We systematically study all reasonable combinations of these degrees of freedom in the coming sections of this paper. A very brief summary of our primary findings is presented in Table \ref{tableone}.

\section{Spin $s=0$ or $s={1\over2}$}\label{Singlets}

We begin with a discussion of particles that are gauge singlets, before moving onto the more interesting case of charged particles.

\subsection{Scalars}

Let us begin with a discussion of a set of singlet scalars, that we denote $\X_I$. Its Lagrangian takes the form (units $\hbar=c=1$, signature $+---$)
\beq
\Delta\mathcal{L} ={1\over2}(\partial \X_I)^2-{1\over2}m_I^2 \X_I^2+\ldots,
\eeq
where the dots represent interactions. If we do not introduce any further internal symmetry, then it is well known that the scalar masses $m_I$ are left unprotected from interactions. Generic interactions among the scalars themselves, or with other particles, will ordinarily lead to the masses $m_I$ becoming very large. A natural value would be towards the GUT scale, or even higher. Since we are searching for models that do not exhibit any fine-tuning, then we are led to assume that indeed $m_I$ is very large. The usual way out of this is to endow the scalars with a global (approximate) symmetry. Namely, a shift symmetry $\X_I\to \X_I+\X_0$, where $\X_0$ is some constant. This forbids a mass term at all orders in perturbation theory. One may then endow $\X_I$ with a mass by appealing to non-perturbative effects that may break the shift symmetry, leading to a small non-zero mass for $\X_I$. This in fact is the situation for axions, which are pseudoscalars associated with some spontaneously broken PQ symmetry \cite{PQ,WeinbergAxion,WilczekAxion}. Having a non-perturbative origin, such a mass is therefore naturally exponentially small.  Such light (but not massless) scalars are interesting dark matter candidates, and indeed there is a large literature on this possibility (e.g., see Refs.~\cite{Preskill:1982cy,Abbott:1982af,Dine:1982ah,Kim:2008hd,Arvanitaki:2009fg}). However, this construction explicitly appeals to the existence of some new global (PQ) symmetry, which is something we are not exploring in this paper; our philosophy is to only impose relativity and quantum mechanics  and nothing else. From the low energy point of view, such a PQ symmetry is arbitrary and is not required by any theoretical consistency arguments. Nevertheless, there are good reasons to think that axions may exist. In the context of string theory, there are string-axions associated with cycles of the compact dimensions. In some contexts, the associated PQ symmetry is in some sense a 4-dimensional relic of the higher dimensional space-time symmetry. On the other hand, achieving sufficient PQ accuracy for the QCD-axion is known to require special model building in the context of quantum gravity \cite{Kamionkowski:1992mf}. 

In any case, we shall not explore that further here. Instead we will focus on theories whose structure is entirely understood within the framework of the low energy effective theory. Hence, we will not introduce any additional arbitrary symmetries, leaving the scalars naturally very heavy.

Since these scalars are naturally very heavy in our framework, their stability is a serious issue. For example, one can readily couple such particles to the Higgs $H$ and other particles in the Standard Model such as the photon $F_{\mu\nu}$, as follows
\beq
\Delta{\mathcal{L}} = \sum_I \X_I \! \left(\mu_I \, H^\dagger H +\gamma_I (F_{\mu\nu}F^{\mu\nu}+F_{\mu\nu}\tilde{F}^{\mu\nu})\right)+\ldots,
\eeq
(the last term is for the case in which $\X_I$ is a pseudoscalar). This will lead to rapid decay of the scalars, with decay rates $\Gamma\sim\mu_I^2/m_I$ or $\Gamma\sim m_I^3/\gamma_I^2$, unless we assume that the couplings $\mu_I,\,\gamma_I$ are exceedingly small, or we assume that the scalar mass is tiny, which is not in accord with our overarching philosophy here. This means that $\X_I$ cannot be a dark matter candidate. The usual idea in the literature to avoid this rapid decay is to appeal to another type of symmetry: a discrete $\mathbb{Z}_2$ symmetry, namely $\X_I\to -\X_I$ (or similar) \cite{Cline:2013gha,Bernal:2015xba}. This symmetry forbids such operators and provides stability of the scalars. However, again we are only interested in relativity and quantum mechanics, so we do not introduce such an ad hoc symmetry. Instead, a consistent way to avoid these operators is to minimally couple such scalars to some spin 1 particles. We will explore this later in Section \ref{ZeroOne}.

\subsection{Fermions}

Another important scenario for the dark matter is that of singlet fermions. This exhibits some similar behavior to that of scalars, with some technical differences that we now describe. For a set of fermions $\F_I$, the Lagrangian is
\beq
\Delta\mathcal{L} = \sum_I \left(i\bar{\F}_I\gamma^\mu\partial_\mu \F_I - m_I\bar{\F}_I \F_I\right)+\ldots,
\eeq
for Dirac fermions, or similar for Majorana, where again the dots represent interactions. As is well known, unlike the case of scalars above, the mass here is protected against interactions due to an emergent chiral symmetry in the limit $m_I\to 0$. For this reason it is technically natural for the masses of fermions to be small. However, again we have no evidence that nature selects this (approximate) chiral symmetry. So we will not appeal to this unneeded symmetry, and instead assume the fermions are heavy; with a mass towards to the GUT scale, or so.

With large masses, there is again a potential problem with stability. Since the fermion is a gauge singlet, there is nothing preventing it from coupling to the Standard Model Higgs $H$ and lepton doublets $L_i$ as follows
\beq
\Delta\mathcal{L}=\sum_{I,i}y_{iI}\,H\,\bar L_i \, \F_I + h.c.,
\eeq
effectively rendering it a type of sterile neutrino. Because of this coupling, the dark matter particle may now decay. In the most natural scenario, this singlet fermion should be very heavy and therefore this process allows for a tree-level decay of $\F$ into Higgs and leptons, followed by later decays into lighter particles. A rough estimate of this decay is $\Gamma\sim y_{iI}^2\,m_I$, which ensures that it will decay rapidly, unless the $y_{iI}$ are extraordinarily small. 

For completeness we can also consider a less natural scenario, in which the fermions are sufficiently light that their decay into Higgs and leptons is not kinematically allowed on-shell. In this case, it is useful to expand the Higgs around its vev, which reveals that the new singlet fermion mixes with neutrinos. So it can oscillate into a neutrino and then through loop diagrams decay into a neutrino and photon.  The rate for this process was estimated in Ref.~\cite{Bulbul:2014sua} to be $\Gamma\sim y_{iI}^2\,m_I^5/M_W^4$. For $\mathcal{O}(1)$ Yukawa couplings, this can only live longer than the present age of the universe for $M_I\lesssim300$\,eV. This motivates many sterile neutrino dark matter models. However, this type of model will not be our focus, since it appeals to an extremely good chiral symmetry that is not demanded by any known fundamental principles.

\section{Dark vs Visible Sector Production}\label{Production}

The above singlet models naturally lead to rapidly decaying particles, and so cannot be the dark matter. In the coming sections we will find stable dark matter candidates when one or more spin $1$ particles are included (plus other particles). The crucial issue then is their abundance. To address this we need to specify its production mechanism, which we discuss briefly here.

\subsection{Coupling to Inflaton}

Our expectation is that the dark matter particle couples to all particles, with interactions only restricted by relativity and quantum mechanics. In particular, there is no reason to expect that it will not couple to the inflaton $\phi$.  Once these couplings are postulated, a thermal history follows that sets the abundance of each sector (also see Refs.~\cite{Feng:2008mu,Adshead:2016xxj,Hardy:2017wkr}).  It is this basic reasoning that allows us to use naturalness to heavily constrain many dark matter models.  However, the precise dynamics involves a variety of processes, and so we will need to compare them all to arrive at the actual thermal history. 

As mentioned above, we will always have at least one spin 1 particle in the dark sector. For purposes of illustration, it suffices to focus on one (unconfined) dark spin 1 particle that the inflaton couples to. We expect the inflaton to be a gauge singlet, nevertheless it can couple to the dark sector gauge field $X$ through its kinetic term $X_{\mu\nu}$ via a dimension 5 operator. Later we will study chiral fermions, whose kinetic term can couple to the inflaton as well, leading to a similar analysis. 

\begin{centering}
	\begin{figure*}[t]
		\centering
		\includegraphics[width=14cm]{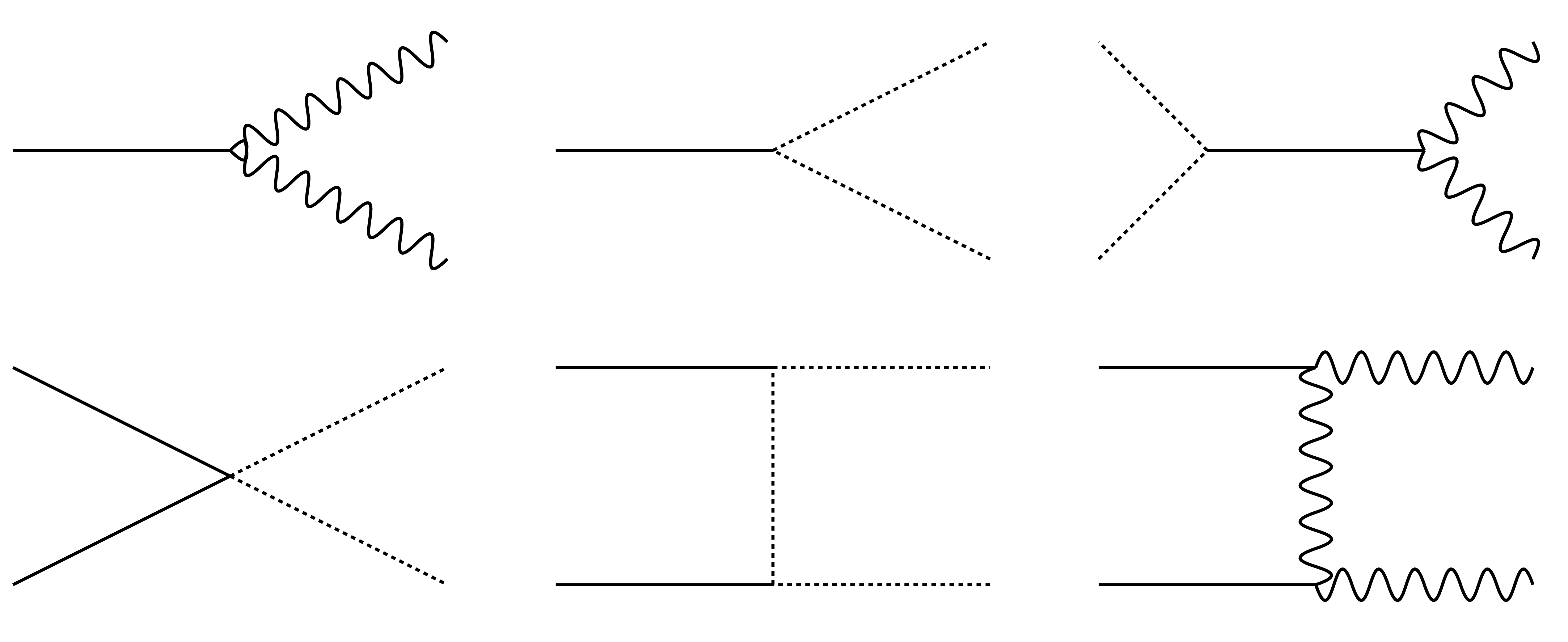}
		\caption{Some of the important processes during reheating.  The solid lines are the inflaton, the dotted lines are the Standard Model Higgs, and the wiggly lines are a spin 1 particle.}
		\label{reheatdiagrams}
	\end{figure*}
\end{centering}

If the inflaton is a real scalar, we also expect it to couple to the Standard Model Higgs $\higgs$ (with associated doublet $H$) through renormalizable operators. Let's write the Lagrangian as
\beq
\Delta\mathcal{L}=\frac{\phi}{\Lam}(X_{\mu\nu}X^{\mu\nu}+X_{\mu\nu}\tilde{X}^{\mu\nu})-\lambda\,\phi^2H^\dagger H+\mu\,\phi\, H^\dagger H+\frac{\phi}{\Lamv}F_{\mu\nu}\tilde{F}^{\mu\nu}+...,
\eeq
We are also allowing for the possibility that the inflaton is a pseudoscalar, which forbids the $\phi\,H^\dagger H$ term, but allows the 2nd, 3rd, and 5th terms. For multiple inflatons, we could have an admixture of both scalar and a pseudoscalar couplings. However, in the simplest case of a single inflaton, as we focus on here, it must be one or the other. These interactions will lead to the processes depicted in Fig. \ref{reheatdiagrams}. Naturalness places bounds on the values of the couplings, as loops involving internal Higgs fields renormalize the mass of the inflaton through logarithmic running, potentially spoiling inflation. For simple inflation models (e.g., $m_\phi^2\phi^2$ chaotic inflation), these bounds are $\lambda\lesssim10^{-5}$ and $\mu\lesssim m_\phi$, where $m_\phi$ is the inflaton mass. A potentially even tighter naturalness bound can be applied to $\lambda$, namely $\lambda\,\UV^2/(4\pi)^2<m_\phi^2$, where $\UV$ is the UV cutoff of the theory. These requirements on $\lambda$ and $\mu$ do not seem especially natural, but we are prepared to accept that the inflationary sector may need to be fine-tuned in order for inflation to persist for a significant number of e-foldings to obtain a large universe. 

Furthermore, a naturalness bound on the dimension 5 couplings $\Lam$, $\Lamv$ is $\Lambda_{UV}^4/((4\pi)^2\Lam^2)<m_\phi^2$. Indeed for a consistent effective theory, we expect $\Lam$ to be very large compared to typical particle mass scales $\Lam\gg m_{\phi}$, etc, and of course we demand $\Lam\gtrsim\Lambda_{UV}$. In fact we will have in mind that at most $\Lam\sim\mpl$, or perhaps an order of magnitude or two smaller. Note that for $\Lam\ll\mpl$, preheating into the dark matter sector will generically occur.  However, this is not very important for our analysis, since the dark sector produced through this process will initially redshift like radiation, and will fail to deplete the inflaton density entirely. In fact after a few e-folds it will be a subdominant component in the universe, assuming the inflaton is heavy. Then standard perturbative reheating will occur.

\subsection{Inflaton Decay}

Now that our expectations for these couplings have been set, the next thing to compute is the production of the dark matter and visible (Standard Model) sectors through these interactions. The first diagram in Fig. \ref{reheatdiagrams} will lead to a decay rate
\beq
\Gamma(\phi\rightarrow XX)\sim \frac{m_\phi^3}{\Lam^2}.
\eeq
In the case in which the dark sector is a significant fraction of the energy density of the universe, one can use $\Gamma\sim H\sim (\Td^{reh})^2/\mpl$ to obtain a dark sector reheat temperature of $\Td^{reh}\sim m_\phi^{3/2}\mpl^{1/2}/\Lam$.  

This can be compared to the decay into the visible sector's Higgs, given by the second diagram in Fig. \ref{reheatdiagrams} for real scalar inflaton, or first diagram for pseudoscalar inflaton, which yields
\beq
\Gamma(\phi\rightarrow \higgs\higgs)\sim\frac{\mu^2}{m_\phi}\,\,\,(\mbox{real}),\hspace{0.8cm}
\Gamma(\phi\rightarrow \higgs\higgs)\sim\frac{m_\phi^3}{\Lamv^2}\,\,\,(\mbox{pseudo}).
\eeq
In the case in which the visible sector is a significant fraction of the energy density of the universe, one can use $\Gamma\sim H\sim (\Tvs^{reh})^2/\mpl$ to obtain a visible sector reheat temperature given by the expression $\Tvs^{reh}\sim \mu\sqrt{\mpl/m_\phi}$ (real) or $\Tvs^{reh}\sim m_\phi^{3/2}\mpl^{1/2}/\Lamv$ (pseudo).

Once the inflaton has decayed, the ratio of the energy densities of the dark to visible sectors is simply the inflaton's decay branching ratio, which in the above model is
\beq
\frac{\rhod}{\rhovs}\sim\frac{m_\phi^4}{\Lam^2\,\mu^2}\,\,\,(\mbox{real}),\hspace{0.8cm}
\frac{\rhod}{\rhovs}\sim\frac{\Lamv^2}{\Lam^2}\,\,\,(\mbox{pseudo}).
\eeq
For real scalar inflaton, if we take $\mu$ towards its maximum natural value of $\mu\sim m_\phi$, and recall that $\Lam\gg m_\phi$ for a consistent effective theory, then this is a relatively small abundance in the dark sector. While for pseudoscalar inflaton, they may be comparable. However, we will find that further interactions may be important (especially in the real scalar case) and can significantly alter the relative abundance.

\subsection{Inflaton Mediation}

Without any direct renormalizable couplings between the two sectors (this will be relevant to some later models, whose charge assignments will forbid this), the only clear way for there to be significant interactions between the two sectors is if it is mediated by inflaton exchange. This is only significant in the real scalar inflaton case, which we focus on here. The cross section for this process is given through the third diagram of Fig. \ref{reheatdiagrams} to be
\beq
\sigma(\higgs\higgs\leftrightarrow XX)\sim \left\{
\begin{array}{cc}
{\mu^2\over\Lam^2\,T^2}, & m_\phi\ll T\\
{\mu^2\,T^2\over\Lam^2\,m_\phi^4}, & m_\phi\gg T
\end{array} \right.
\,,
\eeq
where we assume that the Higgs is relativistic (and $X$ is massless) at these high temperatures. This leads to an annihilation rate, relative to Hubble $H$, of
\beq
\frac{\Gamma(\higgs\higgs\leftrightarrow XX)}{H}\sim \left\{
\begin{array}{cc}
\frac{\mu^2\mpl}{\Lam^2T}, & m_\phi\ll T\\
\frac{\mu^2\mpl T^3}{\Lam^2m_\phi^4}, & m_\phi\gg T
\end{array} \right.
\, .
\eeq
This peaks at the crossover temperature $T\sim m_\phi$ to be $\Gamma(\higgs\higgs\leftrightarrow XX)/H\sim \mu^2\mpl/(\Lam^2m_\phi)$. If we take $\mu$ to be its upper value of $m_\phi$, then in order for this ratio to be larger than 1, we need $\Lam<\sqrt{m_\phi\mpl}$. Naively we might expect $\Lam\sim\mpl$, so this condition would not be satisfied. But lower values of $\Lam$ are possible, so this condition might be satisfied (but it is very unlikely to be satisfied in pseudoscalar case). If so, then the two sectors would thermalize with one another, giving $\Td=\Tvs$, until they later decouple.

\subsection{Inflaton Annihilation}\label{InflatonAnn}

If, on the other hand, $\Lam>\sqrt{m_\phi\mpl}$ then the two sectors will not thermalize with one another, but it is still possible for the inflaton to become thermalized with the Standard Model through the fourth (pseudo) and fifth (real) diagrams of Fig. \ref{reheatdiagrams}.  This rate is given by 
\begin{equation}
\frac{\Gamma(\phi\phi\leftrightarrow \higgs\higgs)}{H}\sim \left\{
\begin{array}{lll}
\frac{\mu^4M_{pl}}{T^5}\,(\mbox{real}) & {\lambda^2\mpl\over T} \,(\mbox{pseudo}) & m_\phi\ll T\\
\frac{\mu^4m_\phi^{3/2}M_{pl}}{T^{13/2}}e^{-m_\phi/T}(\mbox{real}) & \frac{\lambda^2m_\phi^{3/2}M_{pl}}{T^{5/2}}e^{-m_\phi/T}(\mbox{pseudo}) & m_\phi\gg T
\end{array} \right.
\, .
\end{equation}
For real scalar inflaton, this quantity can easily be large; for $T\sim m_\phi$ and $\mu\sim m_\phi$, we have $\Gamma(\phi\phi\leftrightarrow\higgs\higgs)/H\sim\mpl/m_\phi\gg 1$. Hence the inflaton and visible sectors will achieve thermal equilibrium. At this time $T\sim m_\phi$, the energy density for each is $\rho_\phi\sim\rhovs\sim m_\phi^4$. While for pseudoscalar inflaton, the maximum is $\Gamma(\phi\phi\leftrightarrow\higgs\higgs)/H\sim\lambda^2\mpl/m_\phi$, which is expected to be $\ll1$, since we expect $\lambda\lesssim10^{-5}$. So in the pseudoscalar case, equilibrium would not be achieved.

For completeness we include the interaction between inflaton and the dark matter sector given by the last diagram in Fig. \ref{reheatdiagrams}:
\begin{equation}
\frac{\Gamma(\phi\phi\leftrightarrow XX)}{H}\sim \left\{
\begin{array}{ll}
\frac{M_{pl}T^3}{\Lam^4} & m_\phi\ll T\\
\frac{m_\phi^{3/2}M_{pl}T^{3/2}}{\Lam^4}e^{-m_\phi/T} & m_\phi\gg T
\end{array} \right.
\, .
\end{equation}
If we take $T\sim m_\phi$, then we have $\Gamma(\phi\phi\leftrightarrow XX)/H\sim\mpl \,m_\phi^3/\Lam^4$, which is likely to be small, and so the dark sector will likely not be in direct thermal contact with the inflaton.

For the real scalar inflaton case, the inflaton gas generated in this way will then decay, creating a dark sector density at the time of $T\sim m_\phi$ of
\beq
\frac{\rhod}{\rho_\phi}\sim\frac{\Gamma(\phi\rightarrow XX)}{H}\sim\frac{M_{pl}\,m_\phi}{\Lam^2}.
\label{rhodarkrhophi}\eeq
Raising this to the $3/4$ power then gives an estimate of the ratio of number densities. If there are no further number changing processes in the dark sector, then, since the subsequent evolution of the two matter sectors is essentially analogous, aside from baryogenesis, the final abundances would be given by 
\beq
{\Omega_{dark}\over\Omega_{B}}\sim\left(\frac{M_{pl}\,m_\phi}{\Lam^2}\right)^{\!3/4}\frac{m_{DM}}{\eta\, m_p},
\eeq
where $m_p$ is the proton mass and $\eta\sim10^{-9}$ is the baryon-to-photon ratio.  For this ratio to reproduce the observed abundance $\Omega_{dark}/\Omega_B\approx 5$, a dark matter mass of 
\beq
m_{DM}\sim\mbox{MeV} \left(\Lam\over M_{pl}\right)^{\!3/2}\!\left(10^{13}\text{GeV}\over m_\phi\right)^{\!3/4}
\eeq
would be required (we included an extra factor of 10 here due to the fact that eq.~(\ref{rhodarkrhophi}) ignores some factors of $\sim1/(8\pi)$, etc).  This is a very light dark matter mass, which will have a correspondingly large cross section. In any case, our focus is not on such unnaturally small masses. Therefore, the assumption of no further number changing processes in the dark sector needs to be reanalyzed. In fact in the coming sections, we will explore these issues in detail.

\subsection{Parameterization}\label{Parameterization}

Given the above possibilities, let us parameterize the ratio of the temperature of the visible sector $\Tvs$ and the temperature of dark sector $\Td$ just after reheating has occurred in each sector as
\beq
\Td = \tr\, \Tvs. 
\eeq
The value of $\tr$ is determined by the following three basic scenarios: 
\begin{itemize}
\item The two sectors were essentially always decoupled and never thermalized with one another. Instead their production is basically just given by the inflaton decay’s branching ratio, i.e.,
\beq
{\rhod\over\rhovs} = {\Gamma(\phi \to \mbox{DS})\over\Gamma(\phi\to \mbox{VS})},
\eeq
(where ``DS" means dark sector and ``VS" means visible sector).
We then use $\rho\sim g\, T^4$ (where $g$ is the number of degrees of freedom), to obtain the ratio of temperatures as
\beq
\tr = \left({\gvs\,\Gamma(\phi \to \mbox{DS})\over\gd\,\Gamma(\phi\to \mbox{VS})}\right)^{\!1/4}.
\eeq 
\item The two sectors have some significant coupling to each other (whether it is via inflaton mediation or via other particles) and thermalized together at early times, leading to 
\beq
\tr=1. 
\eeq
\item The inflaton thermalizes with the visible sector, but the dark sector remains decoupled, and is only populated via decays. Then the ratio of temperatures is 
\beq
\tr \sim \left({\mpl\,\Gamma(\phi \to \mbox{DS})\over m_\phi^2}\right)^{\!1/4},
\eeq 
(where we suppress the dependence on the number of degrees of freedom here for simplicity).
\end{itemize}
Together, these scenarios potentially allow any $\tr$. However, $\tr\ll1$ is arguably preferred because if the inflaton is a real scalar it can decay directly to Standard Model Higgs via the above dimension 3 operator $\phi\, H^\dagger H$, while in some classes of models there are no such renormalizable operators in the hidden sector. While $\tr\sim1$ may be natural if the inflaton is a pseudoscalar, as it then may decay through dimension 5 operators in either sector more democratically. We note that the late time ratio of temperatures can be moderately different from this; see Section \ref{BBN} for more details.

\section{Spin $s=1$}\label{YangMills}

Since the earlier models involving only gauge singlet particles in the dark sector failed to give the observed relic abundance, due to rapid decays, we now study the case of a dark sector with spin 1 particles. As a starting point, we consider a single gauge group, before considering multiple gauge groups in later sections.  

The case in which scalars are charged under the gauge group will be addressed in the next section. If the gauge group is $SU(\Nd)$, for $\Nd>2$, then any fermion coupled to this gauge group will be of the Dirac type to enforce anomaly cancellation, and so there will be no chiral symmetry forbidding a large mass term\footnote{This neglects special scenarios such as that which occurs in $SU(5)$, where the ${\bf 5}$ representation can cancel the ${\bf\bar{10}}$ representation of Weyl spinors.  It is also circumvented in gauge groups whose representations are manifestly (pseudo)real, which is actually every orthogonal, symplectic or exceptional gauge group aside from $SO(2)$, $SO(4)$, $Sp(2)$, $Sp(4)$, and $E_6$ \cite{Bilal:2008qx}.}.  Similar logic may be applied to the case of $SU(2)$ (and $Sp(\Nd)$), which requires there to be an even number of fermions \cite{Witten:1982fp}.  

For the abelian theory, the particles are simply massless dark photons, and evidently cannot be the dark matter. Furthermore, we are not interested in endowing the spin 1 particles with a mass, because then naturalness allows us to put the mass at a high scale, rendering them irrelevant (in addition to the issue of needing a UV completion).

So our focus is on massless spin 1 associated with a non-abelian gauge group. We expect it to become strongly coupled in the infrared. The lightest bound states of the theory will be entirely determined by pure Yang-Mills theory, and the (heavy) fermion content can be ignored. The corresponding bound states are dark glueballs. 
There are generic problems with overproduction of these states \cite{Halverson:2016nfq}, which we will go through below.  First, however, we quote a generic bound \cite{Soni:2016gzf}, that glueballs are unstable to decay into two gravitons, with lifetime \beq
t_{glue}\sim 10^{10}\,\mbox{yrs}\left(10^7\,\mbox{GeV}\over m_{glue}\right)^{\!5}.
\eeq
So for $m_{glue}\ll 10^7$\,GeV (which will be quite reasonable), they are sufficiently stable to act as dark matter candidates.

\subsection{Glueball Mass and Abundance}

It is possible to estimate the mass of glueballs, as $m_{glue}\sim\Con$, where $\Con$ is the strong coupling scale of the dark gauge group (in the case of $SU(3)$, a more precise estimate is $m_{glue}\approx7\,\Con$). We can determine the strong coupling scale $\Con$ by dimensional transmutation.  The running of the gauge coupling $\ad$ in the dark sector will be given by the standard formula \cite{Caswell:1974gg} to the two loop level:
\beq
\frac{d\ad}{d\ln\mu} = -b_1\alpha^2-b_2\alpha^3,\label{runrun}
\eeq
where the coefficients $b_1$ and $b_2$ are
\beq
b_1=\frac{1}{2\pi}\left(\frac{11}{3}\Nd-\frac23\nf\right),\,\,\,\,\,\, b_2=\frac{1}{8\pi^2}\left(\frac{34}{3}\Nd^2-\frac{20}{3}\Nd-\frac{\Nd^2-1}{\Nd}\nf\right),
\eeq
with $\Nd$ the number of colors and $\nf$ is the number of fermions.

If the coupling is given by $\aUV$ at some fundamental scale $\UV$, such as $\UV=M_{GUT}$, $\UV=M_{string}$, or $\UV=\mpl$, it will attain the strong coupling value $\alpha_s$ at the scale
\beq
\Lambda=\UV\left(\frac{b_2/b_1+\aUV^{-1}}{b_2/b_1+\alpha_s^{-1}}\right)^{b_2/b_1^2}e^{-1/b_1\left(\aUV^{-1}-\alpha_s^{-1}\right)}.
\eeq
Note that the two loop dependence shows up as a subexponential prefactor, which will alter the one loop value by no more than about an order of magnitude.

\begin{centering}
	\begin{figure*}[t]
		\centering
		\includegraphics[width=7.32cm]{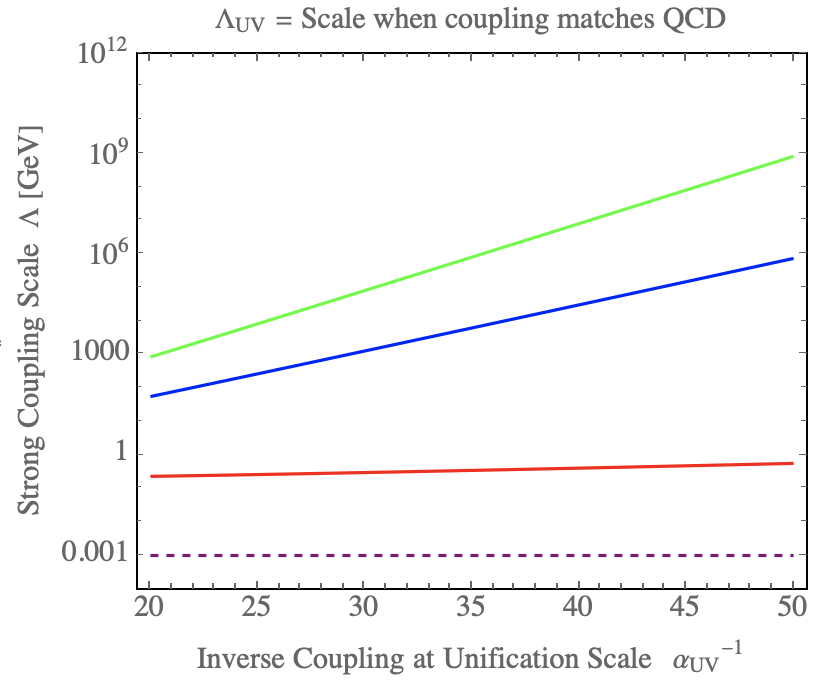}
		\includegraphics[width=9.02cm]{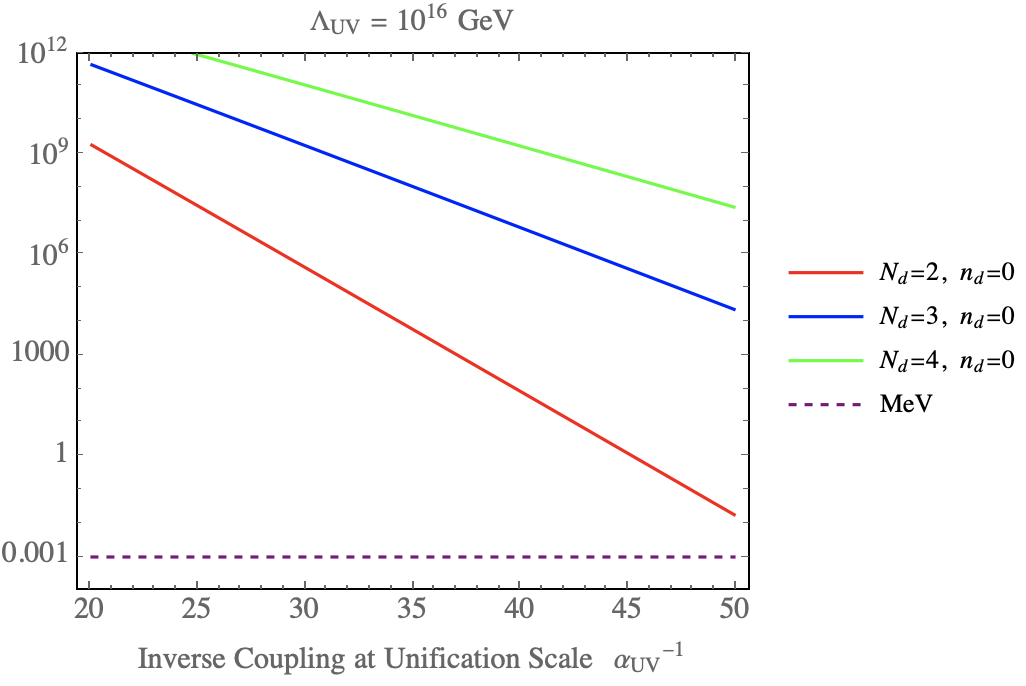}
		\caption{The strong coupling scale of the dark sector $\Con$ (defined as the scale when $\ad=1$) as a function of the (inverse) coupling at the unification scale $\aUV$ for different number of colors $\Nd=2,3,4$ and $\nf=0$ (no dark fermions). Left: The unification scale $\UV$ is defined as the scale at which the coupling matches the Standard Model QCD coupling. Right: The unification scale $\UV$ is defined as $10^{16}\,$GeV.}
		\label{runplot}
	\end{figure*}
\end{centering}

For any particle content in the dark sector, these coefficients can be set, determining the strong coupling scale $\Con$ (which we define as $\alpha_s=1$), which in turn dictates the particle masses for many of the light states in the theory. We would like to determine $\Con$ in terms of the coupling in the context of grand unification. This allows us to specify $\Con$ in terms of the coupling at the unified scale $\aUV$. However, we need to make a choice for the unification scale $\UV=M_{GUT}$. There are at least two ways to specify this: (i) we can define the unification scale as the scale at which $\aUV$ matches the QCD coupling of the Standard Model, (ii) we can simply define $\UV=10^{16}\,$GeV, which is a typical value. These two possibilities are given as the left and hand plots of Fig.~\ref{runplot}, respectively.

After inflation, the dark gluons will acquire a temperature of $\Td\sim\tr\,\Tvs$ (using the parameterization of Section \ref{Parameterization}). They will begin as a relativistic gas, then lock up into glueballs at a temperature $\Td\sim\Con\sim m_{glue}$. From then on they will redshift like matter, leading to an energy density today of $\rhod\sim \Con\,\Td^3$. If we compare this to the baryonic matter energy density today of $\rho_B\sim m_p\,\eta\,T^3$, we obtain a relative abundance of roughly
\beq
{\Omega_{dark}\over\Omega_B}\sim \tr^3 {\Con\over\eta\,m_p}.
\eeq
So if the two sectors had the same or similar temperatures in the early universe after inflation ($\tr\sim1$) then the right abundance for dark matter would require an extremely low strong coupling scale of $\Con\sim\eta\,m_p\sim$\,eV. This would also be a form of hot dark matter and is ruled out. Furthermore, it is rather peculiar to have such an incredibly tiny strong coupling scale. From Fig.~\ref{runplot} we see that this does not fit into any usual unification scenario, which suggest that $\Con$ should be much larger than this. A possible way around this is if $\tr$ were extremely small, for example from the inflaton decaying very slowly into this hidden sector relative to the visible sector. However, such a tiny decay rate seems unusual. From the naturalness arguments of Section \ref{InflatonAnn}, a somewhat small $\tr$, leading to a needed dark mass of $\sim$\,MeV may be plausible (and we have included this for reference in Fig.~\ref{runplot}). Even this is still a very low strong coupling scale, however.

This picture could be altered substantially if number changing interactions are taken into account \cite{Hochberg:2014dra,Boddy:2014yra}, including for example $3\to2$ processes.  In this case, the abundance becomes reduced compared to the above estimates \cite{Bernal:2015ova}. However, such models often make use of light Dirac fermions \cite{Hochberg:2015vrg}, that are unnatural from our more fundamental starting point, and will only be operational for extremely dense environments, $n_{DM}\gtrsim m_{DM}^3$. In any case, such $3\to2$ processes continually produce semi-relativistic particles and hence the dark matter is warm \cite{Pappadopulo:2016pkp}. Such models are ruled out by their smoothing out of small scale structure. Attempts to circumvent this serious problem include adding additional light states or coupling to the Standard Model. However, we are not aware of any model that achieves this within the natural framework we are working.

\subsection{Self-Interaction in Galaxies}\label{SelfInteration}

Note that this scale for the dark glueball sector would lead to a very large cross section \cite{Soni:2016gzf}; the cross section for $2\rightarrow2$ scattering is given by 
\beq
\sigma\sim{1\over \Nd^4\,m_{glue}^2}.
\eeq
If one wishes to have a model of self interacting dark matter that is consistent with galactic observations, the cross section should in fact be 
\beq
{\sigma\over m}\lesssim (0.1-10)\,\mbox{cm}^2/\mbox{g} \sim (0.1-10){1\over(60\,\mbox{MeV})^3}.
\eeq
Beyond this value, there would be noticeable disruption of galaxy collisions \cite{Markevitch:2003at,Harvey:2015hha} and halo characteristics; while if it is of order this value, it may plausibly be an improved fit to the data. If we consider the mass $m_{glue}\sim$\,MeV, then we obtain a cross section  orders of magnitude larger than this galactic bound, which is definitively ruled out by observation.

\section{Spin $s=0$ and $s=1$}\label{ZeroOne}

Since the above models involving pure spin 0 or pure spin $1/2$ (both leading to rapid decays) or pure spin 1 (leading to overabundance) have only shown limited success, in the remaining sections we analyze what happens if we combine these spins.

We first consider the case of a charged scalar $\varphi$. As we are not appealing to any approximate global symmetries, we take the scalar to be very heavy. Then if the gauge is non-abelian then the lightest states are all determined by the strong coupling scale of the theory, as detailed in the previous section, and the heavy scalar is irrelevant. 

\subsection{$U(1)$}

Hence, the only new case to consider is that in which the gauge group is abelian, and hence it does not lead to confinement or the production of any analog of glueballs. The corresponding Lagrangian is well known 
\beq
\mathcal{L}=-{1\over4}F_{\mu\nu}'F^{\mu\nu'}+|\partial_\mu\varphi+i e' A_\mu'\varphi|^2-m_\varphi^2|\varphi|^2+\ldots .
\eeq
The charge $e'$ ensures that such scalars are stable against decay into Standard Model Higgs, etc, as it could in the case of the gauge singlet, and so could potentially play the role of dark matter \cite{Feng:2009mn}. It is possible the scalar is so heavy that it is unable to be produced directly during reheating, in which case its primordial abundance could be negligible (excepting the interesting possibility of freeze-in \cite{Garny:2018grs}). On the other hand, suppose the scalar mass $m_\varphi$ is large, in accord with naturalness, but not so large that in cannot be produced during reheating after inflation; for example, its mass may be $m_\varphi\sim 10^{13}\,$GeV or so. If these particles thermalize, they can then undergo annihilations into dark photons, with a cross section (also see Ref.~\cite{vonHarling:2014kha})
\beq
\sigma(\varphi\varphi\leftrightarrow\gamma'\gamma') \sim {\alpha'^{2}\over m_\varphi^2}.
\eeq
This cross section is extremely small and hence it will cause this process to stop very early, leading to a potentially large relic abundance. If we parameterize the dark sector temperature in the usual way as $\Td=\tr\,\Tvs$, and assuming it began relativistically, before red-shifting, annihilating, and freezing out, the relic abundance of $m_\varphi$ is roughly $\Omega_{dark}\sim \tr\, m_\varphi^2/\alpha'^2/\mbox{TeV}^2$. Since scalar masses $m_\varphi$ are naturally large, this relic abundance is huge, unless the dark sector is extremely cold with a minuscule value of $\tr$.

In these models with scalars charged under a dark $U(1)$, there is nothing forbidding direct couplings to the Standard Model photon and Higgs (sometimes called ``Higgs portal" \cite{Patt:2006fw,Arcadi:2019lka}) via the dimension 4 operators
\beq
\Delta\mathcal{L} = \gamm \,F_{\mu\nu}'F^{\mu\nu} - \lammix\,|\varphi|^2H^\dagger H.
\label{mixing}\eeq
Unless $\gamm$ and $\lammix$ are unreasonably small, this will typically lead to thermalization between the two sectors in the early universe, giving $\tr=1$. This inevitably leads to a huge over abundance.

\section{Spin $s={1\over2}$ and $s=1$}\label{chiral}

Altogether, the above scalar model is not very promising due to the large scalar mass. In this coming sections, we will make use of chiral fermions to provide massless matter content (before confinement) to improve the situation. If the fermions are non-chiral then a Dirac mass term is allowed, and so the fermions would be taken to be very heavy. This would lead to a very similar analysis to the previous section, and will not be repeated here.

Furthermore, we are forced to extend the theory to include multiple gauge groups, so that an initial overabundance of dark matter will partially annihilate into lighter degrees of freedom that will then redshift away, as occurs in the usual freeze-out mechanism.

\subsection{$SU(\Nd)\times U(1)$}

The first possibility is to introduce a dark $U(1)$ alongside the earlier $SU(\Nd)$, but there are some immediate obstacles to this scenario that must be overcome.

The cancellation of the pure $SU(\Nd)$ anomaly still demands that fermions should come in pairs, as above.  The distinction here is that they may in some circumstances have different charges, which prevents a gauge-invariant mass term from being added to the Lagrangian.  The charges are not entirely free, however: anomaly cancellation enforces two additional conditions on the charges of the particles present:
\beq
\sum_i q_i=\sum_iq_i^3=0.
\eeq
For a single species ($\nf=1$), the only solution to these two equations is $q_1=q_2$, so that the two Weyl fermions assemble into a Dirac fermion and hence a large mass term is allowed.  

An interesting setup arises when two species ($\nf=2$) are present, though.  As developed in Refs.~\cite{Harigaya:2016rwr,Co:2016akw}, there are now two branches of solutions to the anomaly cancellation conditions.  The particle content may either be the trivial Dirac fermions, $(\Box,q_1)+(\bar\Box,-q_1)+(\Box,q_2)+(\bar\Box,-q_2)$, or, the nontrivial $(\Box,q_1)+(\bar\Box,-q_2)+(\Box,-q_1)+(\bar\Box,q_2)$.  In this latter case no mass term may be written for the fermions for certain choices of charges $q_i$. On the other hand masses are generated for particles dynamically. There was found to be a possible dark matter candidate as either the associated dark pion (with mass $\sim e'\,\Con$) or dark baryon (with mass $\sim\Con$). The dark condensate that develops carries nonzero dark charge, giving rise to a mass for the dark photon ($\sim \sqrt{\Nd}\,e'\Con$).

Due to the presence of a dark $U(1)$, there is anticipated to be mixing between the Standard Model photon and dark photon, with the form of Eq.~(\ref{mixing}), as discussed at the end of the last section. The bounds on the mixing parameter $\gamm$ are normally very tight. However, since the dark photon is massive in this theory, the bounds on the kinetic mixing are somewhat relaxed \cite{Hochberg:2015vrg}. As elucidated in \cite{Harigaya:2016rwr} can be 
\beq
\gamm\sim10^{-3}\,\,\,\,\, \mbox{for}\,\,\,\,\,m_{\gamma'}\sim10-100\,\mbox{MeV}.
\eeq
This is not an extreme tuning, but does defy the idea of all parameters being $\mathcal{O}(1)$ that we are ultimately searching for to be truly natural.  We will solve this problem in the next section.

Due to the mixing, it is anticipated that the two sectors thermalize with one another in the early universe. Then the relic abundance of dark matter can be determined by the usual freeze-out calculation, whose details will be laid out in more detail in the next Section. Generically one needs a confinement scale $\Con$ that is comparable to $\sim$\,TeV, which is plausible in this framework.

For the scenarios where the dark photon does not acquire a mass, the bounds on charged dark matter coming from the ellipticity of halos is \cite{Feng:2009mn}
\beq
m_d\gtrsim 10^5\alpha_d^{2/3}\,\mbox{GeV}.
\eeq 
However, most important is that there will then be nothing preventing the kinetic mixing with the Standard Model $U(1)$.  As the bounds on this force the coupling to be unnaturally small, this destroys the appeal of this theory.  Alternatively, we note that in Ref.~\cite{Huo:2015nwa} the $U(1)$ was simply taken to be the hypercharge of the Standard Model, circumventing the mixing problem but introducing explicit coupling.

\subsection{$SU(\Nd)\times SU(\Md)$}

\begin{centering}
	\begin{figure*}[t]
		\centering
		\includegraphics[width=7.35cm]{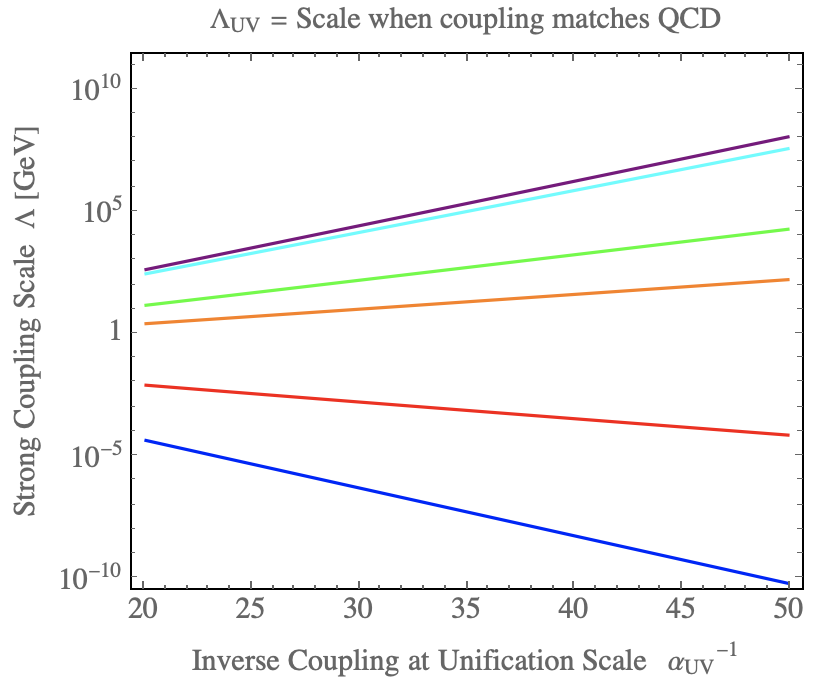}
		\includegraphics[width=8.98cm]{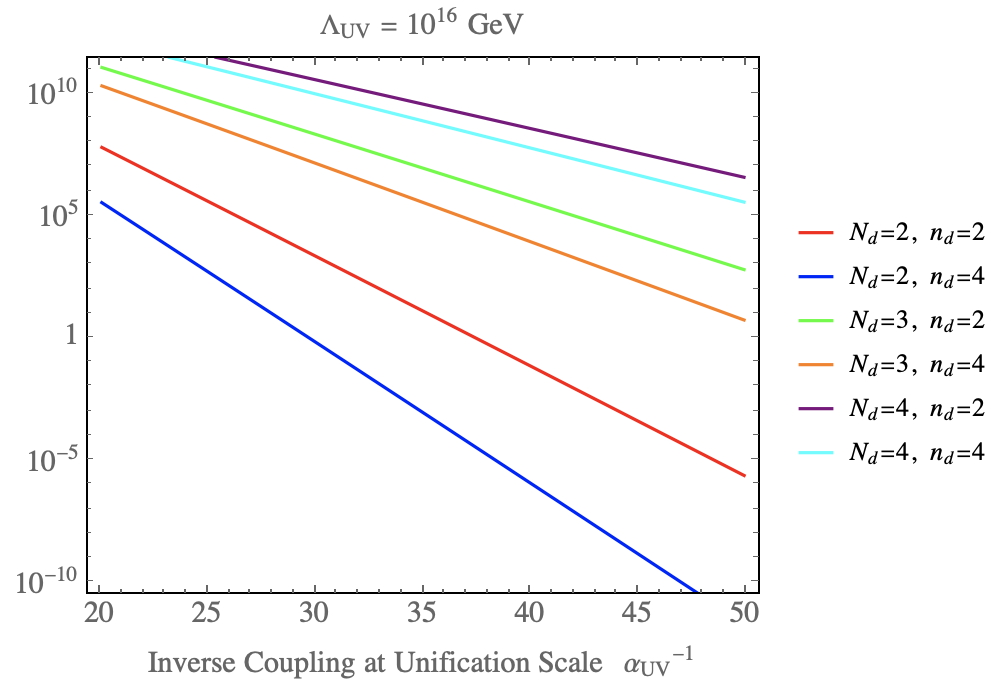}
		\caption{The strong coupling scale of the dark sector $\Con$ (defined as the scale when $\ad=1$) as a function of the (inverse) coupling at the unification scale $\aUV$ for different number of colors $\Nd=2,3,4$ and fermions $\nf=2,4$. Left: The unification scale $\UV$ is defined as the scale at which the coupling matches the Standard Model QCD coupling. Right: The unification scale $\UV$ is defined as $10^{16}\,$GeV.}
		\label{runplot2}
	\end{figure*}
\end{centering}

We turn our attention to $SU(\Nd)\times SU(\Md)$, since a dark non-abelian group will not mix with the Standard Model hypercharge (the logic of \cite{Forestell:2017wov}).  We wish to include chiral fermions in the spectrum of the theory as well, so that they may form light states capable of being produced as radiation in the early universe.  This will alleviate the initial overabundance of dark matter.  

In these theories, anomaly cancellation severely constrains the allowable particle content.  Although mixed anomalies will vanish by the tracelessness of the generators of the special unitary group, the pure $SU(\Nd)$ and $SU(\Md)$ anomalies will in general be nonzero. There are some examples of exceptional cancellations that occur, such as using the {\bf $\bar{5}$} and {\bf 10} representations of $SU(5)$, but these are all special cases that rely on higher representations of the gauge groups.  While these are potentially interesting and deserving of further study, we limit our attention to fundamental representations in this work.

An easier way to satisfy the anomaly cancellation condition is if $\Md=2$, which we will focus on for the remainder of the paper.  In this case, the particle content can be $(\Box,1)+(\bar\Box,1)+(\bar{\Box},\Box)$, which will satisfy all anomaly constraints.  Making the left handed fermions charged under the $SU(2)$ has the added benefit of forbidding a mass term, allowing for naturally light fermions, which we will refer to as dark quarks.  Additionally, if $N_d$ is odd, we must add dark leptons to the theory of the form $(1,\Box)$, due to the Witten anomaly \cite{Witten:1982fp} (any odd number of them could be added, but if they come in pairs, a heavy Dirac mass could be written down, leaving only the odd one in the light spectrum of the theory).  These will be crucial for providing an annihilation channel that induces the freeze-out mechanism in the dark sector.  It remains to determine the low energy spectrum of this theory.  A related theory was considered in \cite{Berryman:2017twh}, however, they included a Higgs to give masses to the gauge bosons. This model is intrinsically unnatural since there is no reason for the Higgs to be light. Instead our focus is on completely Higgs-less theories (or, equivalently, theories in which there is a very heavy Higgs doublet, whose mass is regular and non-tachyonic, and is then irrelevant at low energies).

The dark leptons, if present, will remain massless.  A dark quark condensate will develop, spontaneously breaking the chiral symmetry. This will in fact remove the $SU(2)$ symmetry, giving masses to the associated 3 bosons $W^{1,2,3}$.  In this theory, they will all be degenerate, with masses 
\beq
m_W\sim g\, f_\pi.
\label{mwest}\eeq
For $\Nd\sim3$, one anticipates the pion decay constant will be $f_\pi\sim \Con/3$, as in the Standard Model \cite{Quigg:2009xr}.  For a single generation there will not be any additional massless pions from the spontaneous breaking of chiral symmetry, having all been `eaten' by the $W$ bosons.  These massive $W$ bosons can decay into the massless leptons in the theory. Lastly, the faster running of the $SU(N_d)$ gauge group from Eq. (\ref{runrun}) will naturally cause it to confine, yielding dark baryons comprised of $N$ quarks. There will be significant degeneracy of masses here, so there are no dark quark masses or dark electromagnetic contributions. So, apart from the massless leptons, the lightest stable states are dark analogs of baryons (with no distinction between protons, neutrons, here) with masses
\beq
m_n\sim \mbox{few}\, \Con.
\label{mnest}\eeq
If there are $N_g$ generations, the gauge bosons' masses scale as $\sqrt{N_g}$.  Additionally, there will be $4(N_g^2-1)$ massless pions and $2N_g(4N_g^2-1)/3$ baryons present \cite{Agrawal:1997gf}, leading to potentially more complex dark nuclei than in the visible sector.  Pion masses are computed in \cite{Gasser:1984gg,Donoghue:1994nz,Gao:1996kr}. Dark atoms and molecules were considered in \cite{Cline:2013zca}.

A further class of models includes the full gauge group of the Standard Model $SU(3)\times SU(2)\times U(1)$, with some number of generations of quarks and leptons, but with no light Higgs (as this is unnatural) \cite{Agrawal:1997gf,Quigg:2009xr}. The spectrum of this theory is summarized in Table \ref{tableone}. The inclusion of the additional $U(1)$ is, however, problematic, since now the massless leptons carry a dark (abelian) charge. This means there can be long range forces between these leptons in the late universe, which is potentially in conflict with observation. In fact the running of the dark $\alpha'$ is problematic, as it runs to zero in the infrared, rendering the theory technically trivial. For these reasons, we do not pursue this model further here, and instead we focus on the case in which there is no $U(1)$, and only $SU(\Nd)\times SU(2)$. The special case of $\Nd=3$ will be analyzed in greater detail next.

\section{Dark $SU(3)\times SU(2)$ with 1 Generation of Quarks and Leptons}\label{OneGenModel}

Here we perform the relic abundance calculation for the $SU(3)\times SU(2)$ model, with 1 generation of dark quarks (chiral, charged under $SU(2)$, and charged under $SU(3)$), 1 generation of dark leptons (chiral, charged under $SU(2)$, singlet under $SU(3)$), and no scalars (no Higgs).  

After reheating, this dark sector is taken to have temperature $\Td$, and all the degrees of freedom are unconfined and relativistic. At some lower temperature, the coupling of the dark $SU(3)$ will become strong. In this model, the number of species of dark quarks are $\nf=2$ (as they form a doublet under $SU(2)$), and this strong coupling scale $\Con$ can be read off from the green curve of Fig.~\ref{runplot2}. For temperatures around $\Td\sim\Con$, dark baryons will form; which in this model are dark analogs of protons and neutrons. For $\Td\ll\Con$ the dark baryons will become non-relativistic, and annihilate into lighter particles, depleting their abundance, but eventually freezing out and leaving a relic abundance (of dark baryons and anti-baryons).

In particular, there will be the s-channel annihilation process into two leptons, as well as t and u channel processes into gauge bosons.  The baryon-baryon annihilation cross section is generically impossible to compute directly, as it suffers from strong coupling.  The naive computation into two gauge bosons would have the cross section scale as $\sigma\sim g^4/m_n^2$.  However, this is not the dominant process, as the probability to produce as many gauge bosons as kinematically possible would be the most likely outcome \cite{Ferbel:1969dq}.  A standard estimation of the partial wave cross section, tracing its origins back to Ref.~\cite{Heisenberg:1952}, is \cite{Griest:1989wd}
\beq
\sigma_l v\approx {4\pi(2\ell+1)\over v\, m_n^2}.
\eeq
Additionally, the $2\rightarrow2$ cross section was computed in \cite{Cline:2016nab}, and the dependence on the gauge group used was found to be 
\beq
\sigma\propto {\Nd^4\over(\Nd+1)^2}.
\eeq

In these models the baryon number is stable due to an accidental symmetry as in the Standard Model. Furthermore, since there is only 1 generation of dark quarks and leptons there is no CKM matrix; no baryon asymmetry. Instead the relic abundance is from a symmetric abundance of left over dark baryons and anti-baryons after annihilations freeze-out. Ref.~\cite{Antipin:2015xia} emphasize the appearance of accidental symmetries in related contexts, including an $SU(5)$ grand unified theory. As explored in \cite{Kribs:2016cew}, if dark nuclei are produced during the course of the evolution of the universe, the abundance may be altered, but we do not consider this effect here.

At any rate, the annihilation cross section of a dark baryon with anti-baryon into final state dark leptons at low velocities is
\beq
\langle\sigma v\rangle=\frac{1}{32\pi\,m_n^2}\sum_{\tiny\mbox{final}}|\mathcal{M}(s=4m_n^2)|^2,
\eeq
where the sum is over all final state leptons (it can go into intermediate dark $W$ bosons, which then decay into dark leptons). This can then be used in the relevant Boltzmann equations that determine the abundances
\beq
\frac{d\numd}{dt} = -3H\numd-\langle\sigma v\rangle\left(\numd^2-\numd^{eq}{}^2\right).
\eeq
Here $\numd$ represents the total number density of dark baryons and $\numd^{eq}$ its thermal equilibrium value.  We may define the comoving abundance as $Y_{eq}=x^2K_2(x)$, where $Y$ is number over entropy and $x\equiv \Td/m_n$. Since the temperature of the dark sector may be different than the visible sector at the time of annihilations, as $\Td=\trr\,T$ (where $\trr$ may be mildly different than $\tr$ the ratio right after reheating), we need to be careful. One simple way to account for this is to also define $m\equiv m_n/\trr$. Then the Boltzmann factor that appears in $\numd^{eq}\propto\exp(-m_n/\Td)$, when written in terms of $m$ and $T$, looks canonical $\propto\exp(-m/T)$. We can also estimate the time of freeze-out  as $H(m)\sim m^2/M_{pl}$.  

At early times $\numd\approx \numd^{eq}$, but at late times, the first term of the right hand side of the Boltzmann equation dominates, and the final abundance is approximately $Y_\infty=x_0/(m M_{pl}\langle\sigma v\rangle)$, where $x_0\approx20$ is determined by the time at which the dark matter leaves thermal equilibrium.  The relic abundance is related to this quantity by $\Omega_{dark}=m_n\, Y(\infty)/E_o$, with $E_o\sim3$\,eV, giving the result 
\beq
\Omega_{dark}\approx \trr\,\frac{0.26}{(18\text{ TeV})^2\langle\sigma v\rangle},
\eeq
(we are suppressing a mild logarithmic dependence on the number of degrees of freedom of dark baryons here). The factor of $\trr$ comes from reinstating the proper mass; it accounts for the possibility of different primordial temperatures in the two sectors (see also Ref.~\cite{Feng:2008mu}). Since the presence of dark leptons does not affect the dark matter abundance, this formula holds for any dark matter particle produced by freeze-out from $2\rightarrow2$ interactions. We note that, unlike the glueball models we discussed earlier which can heat up due to $3\to2$ annihilations, that process is forbidden here due to the conserved baryon number.

One may be concerned about the role of dark sector sphalerons. Firstly, we reiterate that there is no (dark) baryon asymmetry in this model for them to wash out. But nevertheless they could play a role in altering the freeze-out abundance. Since there is an SU(2) group, but with no Higgs, we can estimate the sphaleron mass by using the known results for sphalerons in the Standard Model in the $m_H\to\infty$ limit. This is known to give sphaleron masses of $m_{sph}\approx 5.4\,m_W/\alpha_2$, where $\alpha_2=g^2/(4\pi)$ is the SU(2) gauge coupling \cite{Cline:2006ts}. Then using eqs.~(\ref{mwest}--\ref{mnest}), with $m_n\sim4\,\Lambda$ in this model, we have $m_W\sim g\,m_n/12$. Then if we take the Standard Model value of $g\approx 0.65$ at low energies as a guide, we obtain $m_{sph}\sim 9\,m_n$. This means that when the temperature drops below the confinement scale $T<\Lambda$, the sphalerons become highly Boltzmann suppressed $u_{sph}\propto \exp(-m_{sph}/T_{dark})\sim \exp(-9\,m_n/T_{dark})$, much more-so than the dark baryons $u_n\propto \exp(-m_n/T_{dark})$. Hence we expect the sphalerons to in fact be irrelevant to the freeze-out process.

As in \cite{Huo:2015nwa}, if the cross section is scaled from QCD, the correct abundance is produced if $m_n\sim150$\,TeV$/\sqrt{\trr}$. This corresponds to a strong coupling scale $\Con$ roughly 3 times smaller than this,
\beq
\Con \sim 50\,\mbox{TeV}/\sqrt{\trr},
\eeq
for the correct relic abundance. By turning to Fig.~\ref{runplot2}, with $\Nd=3$ and $\nf=2$ (green curve), we see that this is very reasonable. For any $\trr$ that is not incredibly small (in fact $\trr\sim0.1$ may be expected from a real scalar inflaton; see comments in next section), this corresponds to a unification value for $\aUV$ that is beautifully consistent with ideas on grand unification $\aUV\sim 1/35$ (some other choices of $\Nd$ and $\nf$ are promising too). We consider this to be an impressive success of the model.

\section{Big Bang Nucleosynthesis}\label{BBN}

Recall that in the above construction the relic abundance of dark baryons are produced by annihilation into massless dark leptons. This is a form of so-called dark radiation in the universe.  Care must be taken to not spoil the predictions of the early universe by adding additional degrees of freedom during big bang nucleosynthesis (BBN).

The fraction of energy that winds up in dark leptons can be computed by considering the degrees of freedom in each sector, both before and after the major phase transitions that take place.  At sufficiently early times, just after reheating, we parameterize the ratio of temperatures in the two sectors as we did in Section \ref{Parameterization} as $T_{dark} = \tr\, T_{vs}$, where $\tr=1$ if the two sectors begin in thermal equilibrium with each other, and $\tr\neq1$ otherwise. In either case, since the above model does not have renormalizable couplings with the Standard Model, it will decouple from the Standard Model well before confinement of either of the two sectors. Our goal is to then compute the ratio of temperatures of the two sectors at the time of big bang nucleosynthesis and to determine the corresponding change in the effective number of neutrino species.

Let us also allow for the possibility that in the visible sector there are more degrees of freedom than just that of the pure Standard Model $g_{SM}=106.75$. We write the total number of degrees of freedom in visible sector as
\beq
\gvs = \eps\,g_{SM},
\eeq
with $\eps=1$ if the visible sector is purely the Standard Model and $\eps>1$ if there are other particles; these other particles can be new heavy degrees of freedom that directly couple to the Standard Model, such as heavy scalars that couple to the Higgs via $\phi\,H^\dagger H$ or $\phi^2 H^\dagger H$ interactions. If such particles are heavy then they would easily have escaped current detection, but could have been produced in the early universe during reheating. 

Since entropy is conserved independently in each of the two sectors after they decouple from one another, we can write:
\beq
a_*^3 g_{SM*}T_\nu^3 = a^3\gvs \Tvs^3,\,\,\,\,\,a_*^3 g_{dark*}T_{dark*}^3 = a^3g_{dark}T_{dark}^3,
\eeq
where $T_\nu$ is the neutrino temperature after the visible sector has annihilated into just $\gamma,e^-,e^+,\nu$, with degrees of freedom $g_{SM*}=2+{7\over8}(2+2+6)=10.75$, and $T_{dark*}$ is the temperature after the dark sector has undergone confinement. Using the above equations we have
\beq
T_{dark*}={\tr\over\eps^{1/3}}\left(g_{dark}\, g_{SM*}\over g_{dark*}\,g_{SM}\right)^{\!1/3}T_\nu.
\eeq
Then we use the fact the contribution to the energy density is $\Delta\rho\sim \tilde{N}_{dark}\,T^4$, where $\tilde{N}_{dark}$ is the number of light fermionic species in the dark sector. This gives the following contribution to the number of effective neutrino species
\beq
\Delta \Neff={\tr^4\over\eps^{4/3}}\tilde{N}_{dark}\left(g_{dark}\, g_{SM*}\over g_{dark*}\,g_{SM}\right)^{\!4/3}.
\eeq

In the model of the previous section, with $SU(3)\times SU(2)$ and 1 generation of quarks and leptons, we have $g_{dark}=46.5$. After confinement the only light degrees of freedom are a lepton doublet, giving $g_{dark*}=3.5$ and $\tilde{N}_{dark}=2$. Inserting this into the above formula, we have a correction in the number of effective neutrino species during BBN of
\beq
\Delta \Neff\approx 2.9{\tr^4\over\eps^{4/3}}.
\eeq
The current observational bound is $\Delta \Neff\lesssim0.3$ (95\% confidence) \cite{Aghanim:2018eyx}, and this will be improved significantly in upcoming measurements. In order to satisfy this bound we have two basic scenarios: (i) $\tr=1$ for which we need a large number of heavy degrees of freedom in the visible sector; $\eps\gtrsim6$, (ii) $\eps=1$ and the dark sector has a reheat temperature lower than that of the Standard Model; 
\beq
\tr\lesssim0.6.
\eeq
In this framework, with no renormalizable couplings to the dark sector allowed, but renormalizable couplings to the Standard Model Higgs allowed (unless the inflaton is a pseudoscalar), the third scenario outlined in Section \ref{Parameterization} is the most natural. If we push the dimension 5 coupling constant $\Lam$ to the Planck scale $\Lam\sim\mpl$, we have $\tr\sim(m_\phi/\mpl)^{1/4}$, giving $\tr\sim 0.1$ in simple inflation models. This nicely satisfies the above observational bound. If the inflaton is a pseudoscalar, then the first scenario outlined in Section \ref{Parameterization} is the most natural, so comparable temperatures are reasonable, and so while it is not guaranteed, it may well satisfy the above bound too.

Note that the dark particle content we have used represents the best possible attempt at satisfying these bounds, and that adding additional generations or using larger gauge groups would only serve to exacerbate the problem as it would increase $g_{dark}$ hence increasing $\Delta \Neff$.  In this sense, the framework favors the simplest non-trivial extended sectors. Furthermore, we note that models that involve mixing with the Standard Model can, in principle, remove new light degrees of freedom entirely.

Finally, we note that in these models, the scattering cross sections required to achieve the correct relic abundance are so small that there is no problem in satisfying bullet cluster type constraints \cite{Clowe:2006eq,Robertson:2016xjh} on dark matter scattering. This is to be contrasted with the case of glueballs, as discussed earlier in Section \ref{SelfInteration}.

\section{Conclusions}\label{Conclusions}

In this work we began an exploration into models of the dark sector that are truly natural: no small parameters, no fine-tuning, and beyond that, no appeals to any approximate symmetries. By applying these criteria, we found that many constructions fail to give a reasonable dark matter model, for instance singlet scalars or fermions would rapidly decay, charged scalars would have a tiny annihilation cross section leading to overabundance, and similarly pure $SU(\Nd)$ models lead to overabundance of glueballs. 

We found that a kind of minimal model in which all the pieces fit together consistently is a model that involves 11 massless spin 1 particles, organized by gauge groups $SU(3)\times SU(2)$, with a single generation of dark sector quarks and dark sector leptons.  These fermions are chiral and therefore all elementary masses are forbidden; the only masses that are allowed arise from dimensional transmutation. The $\beta$-functions in this model make it entirely reasonable that the confinement scale occurs around $\sim50$\,TeV$/\sqrt{\trr}$ (with $\trr\sim0.1$ arguably preferred by naturalness arguments for real scalar inflatons, or $\trr\sim 1$ for pseudoscalar inflatons), leading to a completely natural ``miracle" in the dark sector with the correct relic abundance of dark baryons. 

A potential observational consequence is that there are massless dark leptons in the model, and so there can be an increase in the effective number of neutrino species $\Neff$ during BBN. However, this problem is avoided by a large number of additional heavy degrees of freedom in the visible sector, or a moderately lower reheat temperature in the dark sector. Both are entirely plausible scenarios, especially the latter, since there are no renormalizable couplings between the inflaton and the dark sector allowed in this construction. Other constructions, with mixing between the dark sector and visible sector are of interest too, which alters this analysis.  Though the model involves a phase transition when the dark gauge group confines, it is expected to be second order \cite{Pisarski:1983ms}, and so it is not expected to yield large cosmological signatures.

Further work is needed to flesh out the details on related models, with other ingredients, as well as to compute the relic abundances with greater precision. Since the models involve strong dynamics, this is a non-trivial step and only order of magnitude estimates have been provided here (although in the specific case of confining group $SU(3)$, we could scale up the QCD results with some precision). It would be very interesting to see to what extent these kinds of models may be embedded in a GUT framework, and/or to elucidate the role of these models within the string landscape. Furthermore, these models indicate that the effective number of relativistic species should be at least slightly larger than the Standard Model prediction, however the deviation is model dependent. It would be very interesting to see whether this value can be pinned down within these other extended contexts and a sharp prediction for BBN can be made.

Our simple estimates of the post-inflationary era suggest that the temperature of the dark sector may in fact be appreciably smaller (perhaps a factor of 10 or so) than the visible sector (unless one considers pseudoscalar inflatons or models with photon mixing, etc). If true, this would imply that the correction to $\Neff$ is essentially unobservable. Also, the simplest models have tiny scattering cross sections, meaning that they do not lead to any appreciable scattering in the galaxy. Furthermore, since many of the above classes of models have no renormalizable couplings to the Standard Model, and yet still produces a beautiful dark matter candidate, it seriously introduces the possible ``nightmare scenario" in which dark matter will remain extremely difficult, if not impossible, to detect. Although this may seem unfortunate, it is a logically consistent possibility, and has completely natural embeddings within particle physics as seen here. In this case, new future search techniques would be appropriate.

\section*{Acknowledgments}
We thank Gary Goldstein and Fuminobu Takahashi for useful discussions.
MPH is supported in part by National Science Foundation grant PHY-1720332.



\begin{thebibliography}{99}

\bibitem{Agrawal:1997gf} 
  V.~Agrawal, S.~M.~Barr, J.~F.~Donoghue and D.~Seckel,
  ``Viable range of the mass scale of the standard model,''
  Phys.\ Rev.\ D {\bf 57}, 5480 (1998)
  [hep-ph/9707380].
  
\bibitem{Weinberg:1987dv} 
  S.~Weinberg,
  ``Anthropic Bound on the Cosmological Constant,''
  Phys.\ Rev.\ Lett.\  {\bf 59}, 2607 (1987).

\bibitem{Kaplan:2009ag} 
  D.~E.~Kaplan, M.~A.~Luty and K.~M.~Zurek,
  ``Asymmetric Dark Matter,''
  Phys.\ Rev.\ D {\bf 79}, 115016 (2009)
  [arXiv:0901.4117 [hep-ph]].
  \bibitem{Cohen:2010kn} 
  T.~Cohen, D.~J.~Phalen, A.~Pierce and K.~M.~Zurek,
  ``Asymmetric Dark Matter from a GeV Hidden Sector,''
  Phys.\ Rev.\ D {\bf 82}, 056001 (2010)
  [arXiv:1005.1655 [hep-ph]].
  \bibitem{Shelton:2010ta} 
  J.~Shelton and K.~M.~Zurek,
  ``Darkogenesis: A baryon asymmetry from the dark matter sector,''
  Phys.\ Rev.\ D {\bf 82}, 123512 (2010)
  [arXiv:1008.1997 [hep-ph]].
  \bibitem{Cheung:2010gj} 
  C.~Cheung, G.~Elor, L.~J.~Hall and P.~Kumar,
  ``Origins of Hidden Sector Dark Matter I: Cosmology,''
  JHEP {\bf 1103}, 042 (2011)
  [arXiv:1010.0022 [hep-ph]].
\bibitem{Das:2010ts} 
  S.~Das and K.~Sigurdson,
  ``Cosmological Limits on Hidden Sector Dark Matter,''
  Phys.\ Rev.\ D {\bf 85}, 063510 (2012)
  [arXiv:1012.4458 [astro-ph.CO]].
\bibitem{Foot:2014uba} 
  R.~Foot and S.~Vagnozzi,
  ``Dissipative hidden sector dark matter,''
  Phys.\ Rev.\ D {\bf 91}, 023512 (2015)
  [arXiv:1409.7174 [hep-ph]].
  \bibitem{Blinov:2012hq} 
  N.~Blinov, D.~E.~Morrissey, K.~Sigurdson and S.~Tulin,
  ``Dark Matter Antibaryons from a Supersymmetric Hidden Sector,''
  Phys.\ Rev.\ D {\bf 86}, 095021 (2012)
  [arXiv:1206.3304 [hep-ph]].
  \bibitem{Lonsdale:2014wwa} 
  S.~J.~Lonsdale and R.~R.~Volkas,
  ``Grand unified hidden-sector dark matter,''
  Phys.\ Rev.\ D {\bf 90}, no. 8, 083501 (2014)
  Erratum: [Phys.\ Rev.\ D {\bf 91}, no. 12, 129906 (2015)]
  [arXiv:1407.4192 [hep-ph]].
\bibitem{Buckley:2014fba} 
M.~R.~Buckley, D.~Feld and D.~Goncalves,
``Scalar Simplified Models for Dark Matter,''
Phys.\ Rev.\ D {\bf 91}, 015017 (2015)
[arXiv:1410.6497 [hep-ph]].
\bibitem{Elor:2015bho} 
  G.~Elor, N.~L.~Rodd, T.~R.~Slatyer and W.~Xue,
  ``Model-Independent Indirect Detection Constraints on Hidden Sector Dark Matter,''
  JCAP {\bf 1606}, no. 06, 024 (2016)
  [arXiv:1511.08787 [hep-ph]].
\bibitem{Acharya:2016fge} 
  B.~S.~Acharya, S.~A.~R.~Ellis, G.~L.~Kane, B.~D.~Nelson and M.~J.~Perry,
  ``The lightest visible-sector supersymmetric particle is likely to be unstable,''
  Phys.\ Rev.\ Lett.\  {\bf 117}, 181802 (2016)
  [arXiv:1604.05320 [hep-ph]].
\bibitem{Dienes:2016vei} 
  K.~R.~Dienes, F.~Huang, S.~Su and B.~Thomas,
  ``Dynamical Dark Matter from Strongly-Coupled Dark Sectors,''
  Phys.\ Rev.\ D {\bf 95}, no. 4, 043526 (2017)
  [arXiv:1610.04112 [hep-ph]].
\bibitem{Escudero:2017yia} 
  M.~Escudero, S.~J.~Witte and D.~Hooper,
  ``Hidden Sector Dark Matter and the Galactic Center Gamma-Ray Excess: A Closer Look,''
  JCAP {\bf 1711}, no. 11, 042 (2017)
  [arXiv:1709.07002 [hep-ph]].
  \bibitem{Tsao:2017vtn} 
  K.~H.~Tsao,
  ``FIMP Dark Matter Freeze-in Gauge Mediation and Hidden Sector,''
  J.\ Phys.\ G {\bf 45}, no. 7, 075001 (2018)
  [arXiv:1710.06572 [hep-ph]].

\bibitem{Linde:1991km} 
  A.~D.~Linde,
  ``Axions in inflationary cosmology,''
  Phys.\ Lett.\ B {\bf 259}, 38 (1991).
\bibitem{Wilczek:2004cr} 
  F.~Wilczek,
  ``A Model of anthropic reasoning, addressing the dark to ordinary matter coincidence,''
  In *Carr, Bernard (ed.): Universe or multiverse* 151-162
  [hep-ph/0408167].
\bibitem{Tegmark:2005dy} 
  M.~Tegmark, A.~Aguirre, M.~Rees and F.~Wilczek,
  ``Dimensionless constants, cosmology and other dark matters,''
  Phys.\ Rev.\ D {\bf 73}, 023505 (2006)
  [astro-ph/0511774].
 
\bibitem{Hellerman:2005yi} 
  S.~Hellerman and J.~Walcher,
  ``Dark matter and the anthropic principle,''
  Phys.\ Rev.\ D {\bf 72}, 123520 (2005)
  [hep-th/0508161].

\bibitem{Susskind:2003kw} 
  L.~Susskind,
  ``The Anthropic landscape of string theory,''
  In *Carr, Bernard (ed.): Universe or multiverse?* 247-266
  [hep-th/0302219].
 \bibitem{Grana:2005jc} 
  M.~Grana,
  ``Flux compactifications in string theory: A Comprehensive review,''
  Phys.\ Rept.\  {\bf 423}, 91 (2006)
  [hep-th/0509003].
  \bibitem{Douglas:2006es} 
  M.~R.~Douglas and S.~Kachru,
  ``Flux compactification,''
  Rev.\ Mod.\ Phys.\  {\bf 79}, 733 (2007)
  [hep-th/0610102].
  
 \bibitem{tHooft:1979rat} 
  G.~'t Hooft,
  ``Naturalness, chiral symmetry, and spontaneous chiral symmetry breaking,''
  NATO Sci.\ Ser.\ B {\bf 59}, 135 (1980).
  
\bibitem{Weinberg:1964ew} 
  S.~Weinberg,
  ``Photons and Gravitons in S Matrix Theory: Derivation of Charge Conservation and Equality of Gravitational and Inertial Mass,''
  Phys.\ Rev.\  {\bf 135}, B1049 (1964).

  \bibitem{Kallosh:1995hi} 
  R.~Kallosh, A.~D.~Linde, D.~A.~Linde and L.~Susskind,
  ``Gravity and global symmetries,''
  Phys.\ Rev.\ D {\bf 52}, 912 (1995)
  [hep-th/9502069].
  \bibitem{Banks:2010zn} 
  T.~Banks and N.~Seiberg,
  ``Symmetries and Strings in Field Theory and Gravity,''
  Phys.\ Rev.\ D {\bf 83}, 084019 (2011)
  [arXiv:1011.5120 [hep-th]].
  \bibitem{Harlow:2018tng} 
  D.~Harlow and H.~Ooguri,
  ``Symmetries in quantum field theory and quantum gravity,''
  arXiv:1810.05338 [hep-th].

\bibitem{Barr:2007rd} 
  S.~M.~Barr and A.~Khan,
  ``Anthropic tuning of the weak scale and of m(u) / m(d) in two-Higgs-doublet models,''
  Phys.\ Rev.\ D {\bf 76}, 045002 (2007)
  [hep-ph/0703219 [HEP-PH]].

\bibitem{Cata:2014sta} 
  O.~Cata and A.~Ibarra,
  ``Dark Matter Stability without New Symmetries,''
  Phys.\ Rev.\ D {\bf 90}, no. 6, 063509 (2014)
  [arXiv:1404.0432 [hep-ph]].
  \bibitem{Antipin:2015xia} 
  O.~Antipin, M.~Redi, A.~Strumia and E.~Vigiani,
  ``Accidental Composite Dark Matter,''
  JHEP {\bf 1507}, 039 (2015)
  [arXiv:1503.08749 [hep-ph]].
  \bibitem{Mambrini:2015sia} 
  Y.~Mambrini, S.~Profumo and F.~S.~Queiroz,
  ``Dark Matter and Global Symmetries,''
  Phys.\ Lett.\ B {\bf 760}, 807 (2016)
  [arXiv:1508.06635 [hep-ph]].

\bibitem{Hertzberg:2017nzl} 
 M.~P.~Hertzberg and M.~Sandora,
  ``Special Relativity from Soft Gravitons,''
  Phys.\ Rev.\ D {\bf 96}, no. 8, 084048 (2017)
  [arXiv:1704.05071 [hep-th]].

 \bibitem{PQ} R.~D.~Peccei and H.~R.~Quinn, 
 ``CP conservation in the presence of instantons,"
Phys.~Rev.~Lett. 38, 1440 (1977).
\bibitem{WeinbergAxion} S. Weinberg, 
``A new light boson?,"
Phys.~Rev.~Lett. 40, 223 (1978).
\bibitem{WilczekAxion} F. Wilczek, 
``Problem of strong P and T invariance in the Presence of instantons," 
Phys.~Rev.~Lett. 40, 279 (1978). 

  \bibitem{Preskill:1982cy} 
  J.~Preskill, M.~B.~Wise and F.~Wilczek,
  ``Cosmology of the Invisible Axion,''
  Phys.\ Lett.\ B {\bf 120}, 127 (1983).
\bibitem{Abbott:1982af} 
  L.~F.~Abbott and P.~Sikivie,
  ``A Cosmological Bound on the Invisible Axion,''
  Phys.\ Lett.\ B {\bf 120}, 133 (1983).
\bibitem{Dine:1982ah} 
  M.~Dine and W.~Fischler,
  ``The Not So Harmless Axion,''
  Phys.\ Lett.\ B {\bf 120}, 137 (1983).
   \bibitem{Kim:2008hd} 
  J.~E.~Kim and G.~Carosi,
  ``Axions and the Strong CP Problem,''
  Rev.\ Mod.\ Phys.\  {\bf 82}, 557 (2010)
  [arXiv:0807.3125 [hep-ph]].
    \bibitem{Arvanitaki:2009fg} 
  A.~Arvanitaki, S.~Dimopoulos, S.~Dubovsky, N.~Kaloper and J.~March-Russell,
  ``String Axiverse,''
  Phys.\ Rev.\ D {\bf 81}, 123530 (2010)
  [arXiv:0905.4720 [hep-th]].

\bibitem{Kamionkowski:1992mf} 
  M.~Kamionkowski and J.~March-Russell,
  ``Planck scale physics and the Peccei-Quinn mechanism,''
  Phys.\ Lett.\ B {\bf 282}, 137 (1992)
  [hep-th/9202003].
  
\bibitem{Cline:2013gha} 
J.~M.~Cline, K.~Kainulainen, P.~Scott and C.~Weniger,
``Update on scalar singlet dark matter,''
Phys.\ Rev.\ D {\bf 88}, 055025 (2013)
Erratum: [Phys.\ Rev.\ D {\bf 92}, no. 3, 039906 (2015)]
[arXiv:1306.4710 [hep-ph]].
\bibitem{Bernal:2015xba} 
N.~Bernal and X.~Chu,
``$\mathbb {Z}_2$ SIMP Dark Matter,''
JCAP {\bf 1601}, 006 (2016)
[arXiv:1510.08527 [hep-ph]].

\bibitem{Bulbul:2014sua} 
E.~Bulbul, M.~Markevitch, A.~Foster, R.~K.~Smith, M.~Loewenstein and S.~W.~Randall,
``Detection of An Unidentified Emission Line in the Stacked X-ray spectrum of Galaxy Clusters,''
Astrophys.\ J.\  {\bf 789}, 13 (2014)
[arXiv:1402.2301 [astro-ph.CO]].

\bibitem{Feng:2008mu} 
  J.~L.~Feng, H.~Tu and H.~B.~Yu,
  ``Thermal Relics in Hidden Sectors,''
  JCAP {\bf 0810}, 043 (2008)
  [arXiv:0808.2318 [hep-ph]].  
\bibitem{Adshead:2016xxj} 
  P.~Adshead, Y.~Cui and J.~Shelton,
  ``Chilly Dark Sectors and Asymmetric Reheating,''
  JHEP {\bf 1606}, 016 (2016)
  [arXiv:1604.02458 [hep-ph]].
\bibitem{Hardy:2017wkr} 
E.~Hardy and J.~Unwin,
``Symmetric and Asymmetric Reheating,''
JHEP {\bf 1709}, 113 (2017)
[arXiv:1703.07642 [hep-ph]].

\bibitem{Bilal:2008qx} 
A.~Bilal,
``Lectures on Anomalies,''
arXiv:0802.0634 [hep-th].
\bibitem{Witten:1982fp} 
E.~Witten,
``An SU(2) Anomaly,''
Phys.\ Lett.\ B {\bf 117}, 324 (1982)
[Phys.\ Lett.\  {\bf 117B}, 324 (1982)].

\bibitem{Halverson:2016nfq} 
J.~Halverson, B.~D.~Nelson and F.~Ruehle,
``String Theory and the Dark Glueball Problem,''
Phys.\ Rev.\ D {\bf 95}, no. 4, 043527 (2017)
[arXiv:1609.02151 [hep-ph]].
\bibitem{Soni:2016gzf} 
A.~Soni and Y.~Zhang,
``Hidden SU(N) Glueball Dark Matter,''
Phys.\ Rev.\ D {\bf 93}, no. 11, 115025 (2016)
[arXiv:1602.00714 [hep-ph]].

\bibitem{Caswell:1974gg} 
W.~E.~Caswell,
``Asymptotic Behavior of Nonabelian Gauge Theories to Two Loop Order,''
Phys.\ Rev.\ Lett.\  {\bf 33}, 244 (1974).

\bibitem{Hochberg:2014dra} 
Y.~Hochberg, E.~Kuflik, T.~Volansky and J.~G.~Wacker,
``Mechanism for Thermal Relic Dark Matter of Strongly Interacting Massive Particles,''
Phys.\ Rev.\ Lett.\  {\bf 113}, 171301 (2014)
[arXiv:1402.5143 [hep-ph]].
\bibitem{Boddy:2014yra} 
K.~K.~Boddy, J.~L.~Feng, M.~Kaplinghat and T.~M.~P.~Tait,
``Self-Interacting Dark Matter from a Non-Abelian Hidden Sector,''
Phys.\ Rev.\ D {\bf 89}, no. 11, 115017 (2014)
[arXiv:1402.3629 [hep-ph]].
\bibitem{Bernal:2015ova} 
N.~Bernal, X.~Chu, C.~Garcia-Cely, T.~Hambye and B.~Zaldivar,
``Production Regimes for Self-Interacting Dark Matter,''
JCAP {\bf 1603}, no. 03, 018 (2016)
[arXiv:1510.08063 [hep-ph]].

\bibitem{Hochberg:2015vrg} 
Y.~Hochberg, E.~Kuflik and H.~Murayama,
``SIMP Spectroscopy,''
JHEP {\bf 1605}, 090 (2016)
[arXiv:1512.07917 [hep-ph]].
\bibitem{Pappadopulo:2016pkp} 
D.~Pappadopulo, J.~T.~Ruderman and G.~Trevisan,
``Dark matter freeze-out in a nonrelativistic sector,''
Phys.\ Rev.\ D {\bf 94}, no. 3, 035005 (2016)
[arXiv:1602.04219 [hep-ph]].

\bibitem{Markevitch:2003at} 
  M.~Markevitch {\it et al.},
  ``Direct constraints on the dark matter self-interaction cross-section from the merging galaxy cluster 1E0657-56,''
  Astrophys.\ J.\  {\bf 606}, 819 (2004)
  [astro-ph/0309303].
\bibitem{Harvey:2015hha} 
D.~Harvey, R.~Massey, T.~Kitching, A.~Taylor and E.~Tittley,
``The non-gravitational interactions of dark matter in colliding galaxy clusters,''
Science {\bf 347}, 1462 (2015)
[arXiv:1503.07675 [astro-ph.CO]].

\bibitem{Feng:2009mn} 
J.~L.~Feng, M.~Kaplinghat, H.~Tu and H.~B.~Yu,
``Hidden Charged Dark Matter,''
JCAP {\bf 0907}, 004 (2009)
[arXiv:0905.3039 [hep-ph]].

\bibitem{Garny:2018grs} 
  M.~Garny, A.~Palessandro, M.~Sandora and M.~S.~Sloth,
  ``Charged Planckian Interacting Dark Matter,''
  JCAP {\bf 1901}, no. 01, 021 (2019)
  [arXiv:1810.01428 [hep-ph]].

\bibitem{vonHarling:2014kha} 
B.~von Harling and K.~Petraki,
``Bound-state formation for thermal relic dark matter and unitarity,''
JCAP {\bf 1412}, 033 (2014)
[arXiv:1407.7874 [hep-ph]].
  
\bibitem{Patt:2006fw} 
  B.~Patt and F.~Wilczek,
  ``Higgs-field portal into hidden sectors,''
  hep-ph/0605188.
\bibitem{Arcadi:2019lka} 
  G.~Arcadi, A.~Djouadi and M.~Raidal,
  ``Dark Matter through the Higgs portal,''
  arXiv:1903.03616 [hep-ph].
  
\bibitem{Harigaya:2016rwr} 
K.~Harigaya and Y.~Nomura,
``Light Chiral Dark Sector,''
Phys.\ Rev.\ D {\bf 94}, no. 3, 035013 (2016)
[arXiv:1603.03430 [hep-ph]].
\bibitem{Co:2016akw} 
R.~T.~Co, K.~Harigaya and Y.~Nomura,
``Chiral Dark Sector,''
Phys.\ Rev.\ Lett.\  {\bf 118}, no. 10, 101801 (2017)
[arXiv:1610.03848 [hep-ph]].

\bibitem{Huo:2015nwa} 
R.~Huo, S.~Matsumoto, Y.~L.~Sming Tsai and T.~T.~Yanagida,
``A scenario of heavy but visible baryonic dark matter,''
JHEP {\bf 1609}, 162 (2016)
[arXiv:1506.06929 [hep-ph]].

\bibitem{Forestell:2017wov} 
L.~Forestell, D.~E.~Morrissey and K.~Sigurdson,
``Cosmological Bounds on Non-Abelian Dark Forces,''
Phys.\ Rev.\ D {\bf 97}, no. 7, 075029 (2018)
[arXiv:1710.06447 [hep-ph]].

\bibitem{Berryman:2017twh} 
J.~M.~Berryman, A.~de Gouvêa, K.~J.~Kelly and Y.~Zhang,
``Dark Matter and Neutrino Mass from the Smallest Non-Abelian Chiral Dark Sector,''
Phys.\ Rev.\ D {\bf 96}, no. 7, 075010 (2017)
[arXiv:1706.02722 [hep-ph]].

\bibitem{Quigg:2009xr} 
C.~Quigg and R.~Shrock,
``Gedanken Worlds without Higgs: QCD-Induced Electroweak Symmetry Breaking,''
Phys.\ Rev.\ D {\bf 79}, 096002 (2009)
[arXiv:0901.3958 [hep-ph]].

\bibitem{Gasser:1984gg} 
J.~Gasser and H.~Leutwyler,
``Chiral Perturbation Theory: Expansions in the Mass of the Strange Quark,''
Nucl.\ Phys.\ B {\bf 250}, 465 (1985).
\bibitem{Donoghue:1994nz} 
J.~F.~Donoghue,
``Light quark masses and mixing angles,''
hep-ph/9403263.
\bibitem{Gao:1996kr} 
D.~N.~Gao, M.~L.~Yan and B.~A.~Li,
``Dashen's theorem and electromagnetic masses of the mesons,''
hep-ph/9612258.

\bibitem{Cline:2013zca} 
J.~M.~Cline, Z.~Liu, G.~Moore and W.~Xue,
``Composite strongly interacting dark matter,''
Phys.\ Rev.\ D {\bf 90}, no. 1, 015023 (2014)
[arXiv:1312.3325 [hep-ph]].

\bibitem{Ferbel:1969dq} 
T.~Ferbel, J.~A.~Johnson, H.~L.~Kraybill, J.~Sandweiss and H.~D.~Taft,
``Pion production and elastic scattering in anti-proton-proton collisions at 6.94 bev/c,''
Phys.\ Rev.\  {\bf 173}, 1307 (1968).
\bibitem{Heisenberg:1952}
W. Heisenberg,
``Production of mesons as a shock wave problem." Zeit. Phys 133 (1952): 65.
\bibitem{Griest:1989wd} 
K.~Griest and M.~Kamionkowski,
``Unitarity Limits on the Mass and Radius of Dark Matter Particles,''
Phys.\ Rev.\ Lett.\  {\bf 64}, 615 (1990).
\bibitem{Cline:2016nab} 
J.~M.~Cline, W.~Huang and G.~D.~Moore,
``Challenges for models with composite states,''
Phys.\ Rev.\ D {\bf 94}, no. 5, 055029 (2016)
[arXiv:1607.07865 [hep-ph]].

\bibitem{Kribs:2016cew} 
G.~D.~Kribs and E.~T.~Neil,
``Review of strongly-coupled composite dark matter models and lattice simulations,''
Int.\ J.\ Mod.\ Phys.\ A {\bf 31}, no. 22, 1643004 (2016)
[arXiv:1604.04627 [hep-ph]].

\bibitem{Cline:2006ts} 
  J.~M.~Cline,
  ``Baryogenesis,''
  hep-ph/0609145.

\bibitem{Aghanim:2018eyx} 
  N.~Aghanim {\it et al.} [Planck Collaboration],
  ``Planck 2018 results. VI. Cosmological parameters,''
  arXiv:1807.06209 [astro-ph.CO].

\bibitem{Clowe:2006eq} 
  D.~Clowe, M.~Bradac, A.~H.~Gonzalez, M.~Markevitch, S.~W.~Randall, C.~Jones and D.~Zaritsky,
  ``A direct empirical proof of the existence of dark matter,''
  Astrophys.\ J.\  {\bf 648}, L109 (2006)
  [astro-ph/0608407].
\bibitem{Robertson:2016xjh} 
A.~Robertson, R.~Massey and V.~Eke,
``What does the Bullet Cluster tell us about self-interacting dark matter?,''
Mon.\ Not.\ Roy.\ Astron.\ Soc.\  {\bf 465}, no. 1, 569 (2017)
[arXiv:1605.04307 [astro-ph.CO]].

\bibitem{Pisarski:1983ms} 
R.~D.~Pisarski and F.~Wilczek,
``Remarks on the Chiral Phase Transition in Chromodynamics,''
Phys.\ Rev.\ D {\bf 29}, 338 (1984).
\end{thebibliography}
\end{document}